\documentclass[a4paper,11pt]{article}

\usepackage{jcappub} 

\usepackage[T1]{fontenc} 
\usepackage{graphicx}
\graphicspath{{figures/}}
\usepackage[utf8]{inputenc}
\usepackage{amsmath}
\usepackage{amsfonts}
\usepackage{amssymb}
\usepackage{enumitem}
\usepackage{multirow}
\usepackage{orcidlink}
\usepackage{CJKutf8}
\usepackage{color}
\usepackage[capitalise]{cleveref}
\usepackage{acro}
\usepackage{adjustbox}
\usepackage{array}
\usepackage{natbib}
\usepackage{dblfnote}
\DFNalwaysdouble
\usepackage{slashed}
\usepackage{dcolumn}
\usepackage{hyperref}
\usepackage{geometry}
\usepackage{subfig}
\usepackage{overpic}
\usepackage{inputenc}
\usepackage{caption}
\usepackage[theorems,skins]{tcolorbox}
\usepackage{booktabs}
\usepackage{tabularx}
\usepackage{array}
\usepackage{ragged2e}

\newcolumntype{L}[1]{>{\RaggedRight\arraybackslash}p{#1}}
\newcolumntype{Y}{>{\RaggedRight\arraybackslash}X}

\def\[{\left[}
\def\]{\right]}
\def\be{\begin{eqnarray}}
\def\ee{\end{eqnarray}}

\DeclareAcronym{GW}{
  short = GW ,
  long = gravitational wave ,
  short-plural = s
}
\DeclareAcronym{LIGO}{
  short = LIGO ,
  long = Laser Interferometer Gravitational-wave Observatory ,
  short-plural =
}
\DeclareAcronym{LISA}{
  short = LISA ,
  long = Laser Interferometer Space Antenna ,
  short-plural =
}
\DeclareAcronym{SKA}{
  short = SKA ,
  long = Square Kilometre Array ,
  short-plural =
}

\DeclareAcronym{SNR}{
	short = SNR ,
	long = signal-to-noise ratio ,
	short-plural =
}

\DeclareAcronym{PTA}{
	short = PTA ,
	long = pulsar timing array ,
	short-plural =
}

\DeclareAcronym{FLRW}{
  short = FLRW ,
  long = Friedmann-Lemaitre-Robertson-Walker ,
  short-plural =
}

\DeclareAcronym{SIGW}{
	short = SIGW ,
	long = scalar induced gravitational wave ,
	short-plural =  s
}

\DeclareAcronym{PBH}{
	short = PBH ,
	long = primordial black hole ,
	short-plural =  s
}

\DeclareAcronym{SMBHB}{
  short = SMBHB ,
  long = supermassive black hole binary ,
  short-plural = s
}

\DeclareAcronym{1PI}{
	short = 1PI ,
	long = one-particle irreducible  ,
	short-plural =
}

\DeclareAcronym{1PR}{
	short = 1PR ,
	long = one-particle reducible  ,
	short-plural =
}

\DeclareAcronym{KDE}{
  short = KDE ,
  long = kernel density estimator ,
  short-plural = s
}

\DeclareAcronym{BPBHM}{
  short = BPBHM ,
  long = binary primordial black hole merger,
  short-plural = s
}

\DeclareAcronym{CMB}{
	short = CMB ,
	long = cosmic microwave background ,
	short-plural =
}
\DeclareAcronym{DM}{
	short = DM ,
	long = dark matter ,
	short-plural =
}

\DeclareAcronym{BBN}{
	short = BBN ,
	long = Big-Bang nucleosynthesis ,
	short-plural =
}

\DeclareAcronym{LN}{
	short = LN ,
	long = log-normal  ,
	short-plural =
}

\DeclareAcronym{BPL}{
	short = BPL ,
	long = broken power-law ,
	short-plural =
}

\DeclareAcronym{SGWB}{
	short = SGWB ,
	long = stochastic gravitational	wave background ,
	short-plural =  s
}

\DeclareAcronym{LSS}{
	short = LSS ,
	long = large scale structure ,
	short-plural =
}

\DeclareAcronym{RD}{
	short = RD ,
	long = radiation-dominated ,
	short-plural =
}

\DeclareAcronym{PLS}{
	short = PLS ,
	long = power low sensitivity ,
	short-plural =
}

\DeclareAcronym{MAP}{
	short = MAP ,
	long = maximum a posterior ,
	short-plural =
}

\DeclareAcronym{BAO}{
	short = BAO ,
	long = baryon acoustic oscillations ,
	short-plural =
}

\DeclareAcronym{TSIGW}{
	short = TSIGW ,
	long = tensor-scalar induced gravitational wave ,
	short-plural =  s
}

\title{\boldmath Scalar induced gravitational waves as probes of dark QCD}

\author[a]{Wan-Zhe Feng,}
\author[a]{Ao Li,}
\author[b]{Jing-Zhi Zhou}

\affiliation[a]{Center for Joint Quantum Studies and Department of Physics,
School of Science, Tianjin University, Tianjin 300350, China}

\affiliation[b]{ Department of Mathematics and Physics, Huaian University, Meicheng East Road 1, Huaian, Jiangsu 223200, China}

\emailAdd{vicf@tju.edu.cn}
\emailAdd{leo\_6626@tju.edu.cn}
\emailAdd{zhoujingzhi@tju.edu.cn}

\abstract{We investigate scalar induced gravitational waves (SIGWs) as probes of a dark QCD crossover. Motivated by twin Higgs and asymmetric twin baryon dark matter scenarios, we consider a dark QCD sector with a confinement scale approximately 5.5 times the Standard Model (SM) QCD scale. We construct the effective energy and entropy degrees of freedom for the SM supplemented by dark QCD sectors containing either three light dark quark flavors or all six dark quark flavors. The resulting equation of state parameter and sound speed are then used to solve the first-order scalar perturbations and the second-order SIGWs through the SM and dark QCD crossover epochs. For a monochromatic primordial curvature power spectrum, we first demonstrate that the realistic SM thermal history modifies
the SIGW spectrum relative to the idealized radiation-dominated case.
We then show that a dark QCD crossover generates an additional frequency-shifted distortion when the enhanced scalar mode reenters the horizon near the dark confinement scale. This distinctive feature can therefore serve as a characteristic signature of the dark QCD sector. Our results demonstrate that SIGWs provide a complementary cosmological probe of hidden confining sectors, with characteristic spectral features shifted to higher frequencies relative to the SM QCD imprint. The analysis developed in this work can also be extended to other well-motivated theories containing different dark confining sectors.}

\begin{document}

\maketitle
\flushbottom

\section{Introduction}

Gravitational waves provide a unique probe of the thermal history of the early Universe. Unlike electromagnetic signals, a stochastic gravitational wave background can preserve information from epochs long before recombination. In particular, when the equation of state of the cosmic plasma deviates from that of an ideal radiation-dominated background, gravitational wave modes associated with horizon reentry around that epoch can carry characteristic imprints.
The QCD crossover provides an important example. Although it is a smooth crossover rather than a strongly first-order phase transition, it modifies the thermodynamic properties of the plasma, including the effective energy and entropy degrees of freedom, the equation of state parameter $w(T)$, and the sound speed $c_s^2(T)$. These modifications affect the evolution of cosmological perturbations and can therefore leave characteristic imprints on the gravitational wave energy density spectrum.

A precise treatment of these effects requires going beyond the ideal gas approximation. In the Standard Model (SM), the temperature dependence of the effective degrees of freedom receives important contributions from particle thresholds, finite-temperature interactions, non-perturbative QCD dynamics near the crossover, and perturbative QCD at higher temperatures. The importance of these thermodynamic effects for primordial gravitational wave spectra was emphasized in~\cite{Saikawa:2018rcs}, where the SM effective energy and entropy degrees of freedom were reconstructed over a wide temperature range by combining the hadron resonance gas description, lattice QCD results, perturbative QCD, and electroweak corrections. The resulting QCD crossover produces
a smooth but potentially observable distortion in the gravitational wave energy density spectrum associated with modes
entering the horizon near the QCD epoch.

In this work we focus on scalar induced gravitational waves (SIGWs). On small scales, primordial curvature perturbations are not tightly constrained by current cosmological observations. Therefore, if large‑amplitude primordial perturbations exist on these scales, the curvature perturbations couple to second‑order tensor modes through the second‑order cosmological perturbation equations, generating second‑order SIGWs. In addition to the power spectrum of primordial curvature perturbations, SIGWs generated through this mechanism are also affected by the sound speed parameter and the equation of state~\cite{Ananda:2006af,Baumann:2007zm,Domenech:2019quo,Abe:2020sqb,Domenech:2021ztg,Yuan:2021qgz,Balaji:2023ehk,Zhu:2023gmx,Liu:2023pau,Harigaya:2023pmw,Domenech:2024rks}.
Therefore, although gravitational waves do not directly couple to the sound speed of the plasma, SIGWs can serve as an indirect probe of $w(T)$ and $c_s^2(T)$. This mechanism was used in~\cite{Abe:2020sqb} to show that the SM QCD crossover can leave characteristic imprints on SIGWs in the pulsar timing array frequency range.

The same idea can be extended to strongly coupled hidden sectors. A dark QCD sector, consisting of dark quarks and dark gluons charged under a hidden confining gauge group, arises naturally in many extensions of the SM. In particular, twin Higgs constructions provide a well-motivated framework in which a mirror or twin QCD sector appears alongside a solution to the electroweak hierarchy problem.
In asymmetric twin baryon dark matter scenarios, a stable twin baryon can constitute the dark matter. If the visible and twin baryon asymmetries are comparable, the observed ratio $\Omega_{\rm DM}/\Omega_b$ favors a twin baryon mass of a few GeV. This motivates a benchmark with an enhanced dark confinement scale relative to that of SM QCD~\cite{An:2009vq,Farina:2015uea,Feng:2020urb},
\begin{equation}
\Lambda_{\rm dQCD}
\simeq
5.5\,\Lambda_{\rm QCD}\,.
\label{eq:QCDdQCD}
\end{equation}
Such a sector is challenging to probe directly because its states are neutral under the SM gauge group and communicate with the visible sector only through portal interactions or higher-dimensional operators. Nevertheless, its thermal evolution can modify the equation of state of the early Universe and thereby leave characteristic imprints on the SIGW spectrum.
\emph{The method developed here is not restricted to the dark QCD framework considered in this work and can also be applied to other well-motivated theories containing different dark confining sectors.}

The purpose of this paper is to study SIGWs as probes of a dark QCD crossover. We construct the effective energy and entropy degrees of freedom for representative thermal histories containing the SM plus a dark QCD sector. Motivated by mirror dark matter, we consider two benchmark cases: a minimal dark QCD sector with three light dark quark flavors $(u',d',s')$, denoted by ${\rm SM+dQCD}_3$;
and a six-flavor dark QCD sector containing $(u',d',s',c',b',t')$, denoted by ${\rm SM+dQCD}_6$.
In both cases we take the dark confinement scale to be approximately $5.5$ times the SM QCD scale. We model the dark QCD thermodynamics by adapting the SM treatment: a dark hadron resonance gas below the crossover, lattice-inspired results near the crossover, and perturbative QCD above the crossover. This allows us to obtain the temperature-dependent functions $g_{*}(T)$ and $g_{*s}(T)$, and hence the corresponding equation of state and sound speed.

We compute the time-dependent $w(\eta)$ and $c_s^2(\eta)$ induced by the SM and dark QCD thermal histories. On this basis, the evolution of first‑order scalar perturbations and the second‑order SIGWs can be calculated, allowing us to derive the corresponding energy density spectrum of SIGWs. In this paper, we consider the monochromatic primordial power spectrum,
\begin{equation}
    \mathcal{P}_{\zeta}(k)
    = A_{\zeta}\,k_*\,
    \delta(k- k_*)\,,
    \label{eq:mono_spectrum}
\end{equation}
where $k_*$ selects the scale of the enhanced primordial curvature perturbation. When $k_*$ is chosen such that the corresponding scalar mode reenters the horizon near the QCD or dark QCD crossover, the energy density spectrum of SIGWs is maximally sensitive to the corresponding variation of the sound speed. In terms of the physical frequency, the corresponding scale is
\begin{equation}
f_* = \frac{k_*}{2\pi a_0}\,.
\end{equation}
Therefore, the SM QCD and dark QCD crossovers can imprint characteristic features at different frequencies.

This paper is organized as follows. In Section~\ref{Sec:gdQCD}, we describe the dark QCD benchmark motivated by twin Higgs and asymmetric twin baryon dark matter, and construct the effective relativistic degrees of freedom for the SM plus dark QCD thermal bath. In Section~\ref{Sec:SIGW}, we summarize the SIGW formalism used in our calculation. In Section~\ref{Sec:result}, we compute the temperature dependence of $w(T)$ and $c_s^2(T)$ and study the resulting SIGW spectra for the SM only and SM plus dark QCD cases. We conclude in Section~\ref{Sec:Con}. Technical details of the SM effective degrees of freedom, including the QCD trace anomaly, the hadron resonance gas description, perturbative QCD, and heavy quark thresholds, are collected in Appendix~\ref{App:SMg}.
The relation between conformal time and the temperature of the Universe is derived in Appendix~\ref{App:UTtoCT}.

\section{Effective degrees of freedom with a dark QCD sector}\label{Sec:gdQCD}

A dark QCD extension of the SM introduces a new confining non-Abelian gauge group, typically $SU(N)_D$,
with its own dark gluons and dark quarks that are neutral under SM gauge interactions.
The dark QCD sector confines at a scale $\Lambda_{{\rm dQCD}}$,
producing a spectrum of dark hadrons (mesons, baryons, and possibly glueballs),
with the lightest stable state often serving as a natural dark matter candidate.
While the dark QCD sector is hidden,
it can communicate with the SM through portal interactions,
such as the Higgs portal, kinetic mixing (if an additional dark $U(1)$ is present),
or higher-dimensional operators induced by heavy connector states,
thereby leading to rich phenomenology,
from strongly interacting dark matter and exotic collider signatures to distinctive cosmological signals.
Among the various dark QCD extensions, the twin QCD sector arising in twin Higgs constructions stands out as a particularly well-motivated framework, especially in cogenesis scenarios that predict a lightest stable twin baryon with a mass approximately five times that of the proton.

\subsection{Twin Higgs and asymmetric twin baryon dark matter}

Twin Higgs models offer an alternative to supersymmetry for addressing the electroweak hierarchy problem~\cite{Chacko:2005pe, Burdman:2006tz, Cai:2008au, Craig:2014aea, Csaki:2017jby, Serra:2017poj, Xu:2018ofw, Cohen:2018mgv, Xu:2019xuo, Ahmed:2020hiw}.
In these constructions, the Higgs doublet arises as a pseudo-Nambu--Goldstone boson of an approximate global symmetry,
while a discrete mirror symmetry $\mathbb{Z}_2$ relates the SM to a ``twin'' (or mirror) sector.
Radiative corrections to the Higgs potential are thus controlled by the enlarged symmetry structure,
allowing the weak scale to remain natural without introducing new SM colored states at the TeV scale.

A particularly well-motivated cosmological application is \emph{asymmetric} twin dark matter~\cite{An:2009vq,Farina:2015uea,Feng:2020urb},
where the dark matter is a stable twin baryon (e.g., the twin proton) carrying a conserved twin baryon number.
If the baryon asymmetries in the visible and twin sectors are comparable, the observed cosmic coincidence,
$\Omega_{\rm DM}/\Omega_b \simeq \mathcal{O}(5)$~\cite{Ade:2013sjv, Planck:2018vyg},
suggests a dark baryon mass of a few~GeV.
If the asymmetries in the SM and twin sectors are co-generated from the same source and have equal magnitudes,
an asymmetric twin proton dark matter candidate with
\begin{equation}
m_{p'} \simeq 5.5~{\rm GeV}
\end{equation}
can naturally account for the observed cosmic coincidence~\cite{Feng:2020urb}.

In mirror realizations with an approximate $\mathbb{Z}_2$ symmetry,
the twin Yukawa couplings are (to a good approximation) identical to their SM counterparts.
As a result, the twin fermion masses, and in particular the twin quark masses,
scale with the ratio of the twin and SM Higgs vacuum expectation values,
$m_{q'} \simeq (v'/v)m_q$. Therefore, given the experimental constraint $v'/v \gtrsim 3$~\cite{ParticleDataGroup:2024cfk}, one obtains $m_{q'} \gtrsim 3m_q$.
\emph{In this work, we adopt this experimental bound and assume that the dark quark masses are three times those of their corresponding SM quarks.}

Hadron masses, however, are largely set by confinement dynamics.
Therefore, realizing a twin proton mass
$m_{p'}\simeq 5.5~{\rm GeV}$ typically points to an enhanced twin confinement scale,
cf., Eq.~\eqref{eq:QCDdQCD},
which may arise from modest departures from exact $\mathbb{Z}_2$ symmetry in the twin color sector
and/or from threshold effects that modify the running of the twin strong coupling~\cite{Geller:2014kta,Csaki:2015gfd,Barbieri:2015lqa,Ahmed:2017psb}.
This setup provides a simple and predictive framework in which neutral naturalness
and the dark matter--baryon coincidence can be addressed simultaneously.

Consequently, a dark QCD sector with a confinement scale $\Lambda_{\rm dQCD}\sim 5.5\,\Lambda_{\rm QCD}$ and dark quark masses $m_{q'}\sim 3\,m_q$ provides a well-motivated benchmark. Such a sector is generically challenging to probe directly at colliders, since its states are neutral under the SM gauge group.
Nevertheless, if present, it can leave characteristic imprints on the thermal history of the early Universe. In particular, if the dark QCD crossover qualitatively resembles the SM QCD crossover, it modifies the effective relativistic degrees of freedom $g_\ast(T)$ and the thermodynamic evolution of the cosmic plasma. These changes alter the sound speed during the dark QCD and SM QCD crossover epochs, thereby modifying the resulting SIGW energy density spectrum and its detectability in the corresponding frequency range. In this work, we focus primarily on the temperature range encompassing both the dark QCD and SM QCD crossovers.

\subsection{Effective degrees of freedom with a dark QCD sector}

In this subsection, we derive the effective energy and entropy degrees of freedom for two benchmark extensions of the SM, following the procedure summarized in Appendix~\ref{App:SMg}:
\begin{itemize}
  \item the SM supplemented by a minimal dark QCD sector containing three light dark quark flavors (${\rm SM+dQCD}_3$);
  \item the SM plus a dark QCD sector containing all six dark quark flavors (${\rm SM+dQCD}_6$).
\end{itemize}
In Twin Higgs constructions, the leptonic content of the twin sector is highly model dependent.
In particular, constraints on $\Delta N_{\rm eff}$ severely restrict light mirror species,
typically requiring them to annihilate efficiently into the SM before BBN.
We therefore do not model the dynamics of the twin leptonic sector or the dark photon here, and focus on the dark QCD contribution.

\paragraph*{SM plus a minimal dark QCD sector with three light dark quarks (${\rm SM+dQCD}_3$)}

We consider a minimal dark QCD sector containing three light dark quarks $(u^{\prime}$, $d^{\prime}$, $s^{\prime})$, with masses taken to be three times
those of their SM counterparts $u, d, s$, respectively. The dark gauge sector consists of eight massless dark gluons.
We assume an enhanced confinement scale, $\Lambda_{\mathrm{dQCD}} = 5.5 \Lambda_{\mathrm{QCD}}$ ,
and as a representative benchmark, take dark baryon masses to scale accordingly, e.g., the
dark proton mass is $m_{p}^{\prime}\simeq 5.5 m_{p}$.

For simplicity, we do not include the dark Higgs explicitly in the present analysis. The dark Higgs in the setup is sufficiently heavy such that its associated symmetry breaking occurs well before the QCD or dark QCD crossover. Its dynamics are thus expected to decouple from the QCD or dark QCD crossover and should not significantly modify the crossover-induced SIGW signal.

To model the dark sector contribution to the effective relativistic degrees of freedom,
we follow the SM derivation summarized in Appendix~\ref{App:SMg} and divide the calculation into three regimes:
\begin{itemize}
  \item a dark HRG description well below the dark QCD crossover, for $T \lesssim 660~\mathrm{MeV}$;
  \item a lattice description in the vicinity of the crossover, for $660~\mathrm{MeV} \lesssim T \lesssim T_s$;
  \item a pQCD description well above the crossover, for $T \gtrsim T_s $.
\end{itemize}

Since the trace anomaly $\Delta(T)$ and the ratio $p(T)/T^4$ are dimensionless,
their temperature dependence is governed primarily by the dimensionless mass ratios $m/T$.
We therefore approximate the dark QCD contribution from three light flavors by a simple rescaling of the SM light flavor result.
Namely, we shift the characteristic structures in $g_\ast(T)$ to higher temperatures by a factor of $5.5$.

After rescaling the hadron masses accordingly, the HRG contribution to the energy and entropy degrees of freedom for
$T\lesssim 660~\mathrm{MeV}$ can be computed using Eqs.~\eqref{gsr_fitting_function_low} and~\eqref{gss_fitting_function_low}.
In the vicinity of the dark QCD crossover, we obtain $\Delta(T)$ and $p(T)/T^{4}$ from the lattice QCD fitting functions for three light quark flavors over $100~\mathrm{MeV}$--$500~\mathrm{MeV}$ in the SM~\cite{Borsanyi:2016ksw}.
Applying the same temperature rescaling by a factor of $5.5$,
we use these fits to model the dark quark and dark gluon contributions in the range
$660~\mathrm{MeV}\lesssim T \lesssim 2.75~\mathrm{GeV}$.

At high temperatures, we evaluate the dark sector pressure using the pQCD expression in Eq.~\eqref{eq:p_pQCD} with $N_f=3$. To evaluate the running strong coupling $\alpha_s$ at the dark QCD scale, we adopt the same approach as~\cite{Saikawa:2018rcs}, using the four-loop beta function coefficients~\cite{vanRitbergen:1997va}. In practice, we employ the iterative analytic solution of the renormalization group equation given in~\cite{ParticleDataGroup:2018ovx}.

Since the dark sector contains only three light quark flavors, heavy quark mass effects are irrelevant. To match $\Delta(T)$ between the lattice and perturbative regimes in the dark sector,
we consider two choices of the switching temperature, $T_s=2.75~{\rm GeV}$ and $T_s=3~{\rm GeV}$.\footnote{The switching temperatures used in~\cite{Saikawa:2018rcs} for the QCD sector are $500~{\rm MeV}$ and $1~{\rm GeV}$. After rescaling the temperature by a factor of $5.5$, these values correspond to $2.75~{\rm GeV}$ and $5.5~{\rm GeV}$ for the dark QCD sector. Within this rescaled matching window, $\Delta(T)$ is continuous across the matching point by construction. We further require its first derivative to be approximately continuous, in order to ensure a smooth behavior of the pressure and the effective degrees of freedom. We find that this derivative matching condition is better satisfied toward the lower end of the rescaled window, while near the upper end, around $5.5~{\rm GeV}$, the derivative discontinuity becomes more pronounced. We therefore take $T_s=2.75~{\rm GeV}$ and $T_s=3~{\rm GeV}$ as representative switching temperatures on the lower side of the rescaled matching window for the dark QCD.} For each value of $T_s$, we fix the renormalization scale to $\mu=2\pi T$ and vary $q_c(N_f=3)$, cf. Eq.~\eqref{eq:p_QCD_order_g6}, within a conservative range in order to propagate the lattice uncertainty at the matching point.

This procedure yields two representative curves for $\Delta(T)$. After performing the integration in Eq.~\eqref{eq:Deltop}, we obtain the corresponding curves for $p(T)/T^4$, from which we derive the effective degrees of freedom $g_{*}(T)$ and $g_{*s}(T)$. The spread between the two resulting curves is taken as the theoretical uncertainty, while their median is used as the central value in the perturbative QCD regime.

In the $N_f=4$ case, the reference value $q_c(N_f=4)=-3000$ is adopted for the QCD sector~\cite{Saikawa:2018rcs}. By contrast, no corresponding reference value is available for $q_c(N_f=3)$ for the dark QCD sector. We therefore do not employ the alternative matching prescription in which $q_c$ is fixed while the renormalization scale $\mu$ is varied. Instead, we fix $\mu=2\pi T$ and vary $q_c(N_f=3)$ to propagate the matching uncertainty.

\begin{figure}[t]
		\centering
		\includegraphics[width=12cm]{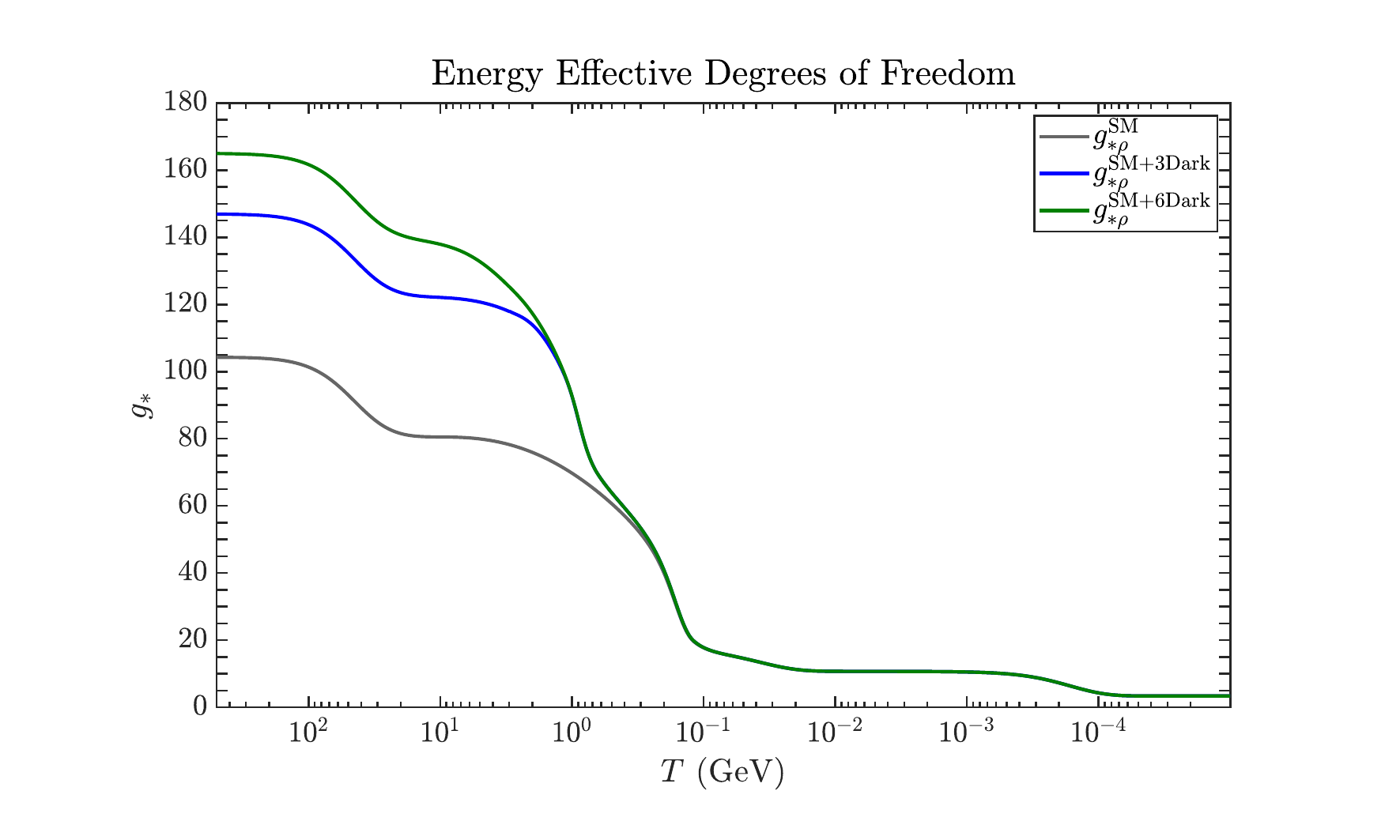}
        \caption{Total effective energy degrees of freedom as functions of temperature for the pure SM (grey),
        the SM plus a minimal dark QCD sector with $(u^\prime,d^\prime,s^\prime)$ (blue),
        and the SM plus a dark QCD sector with all six dark quark flavors (dark green).
        The effective degrees of freedom of the eight dark gluons are included for both ${\rm SM+dQCD}_3$ and ${\rm SM+dQCD}_6$.}
		\label{fig: EdQCD}
\end{figure}

\begin{figure}[t]
		\centering
		\includegraphics[width=12cm]{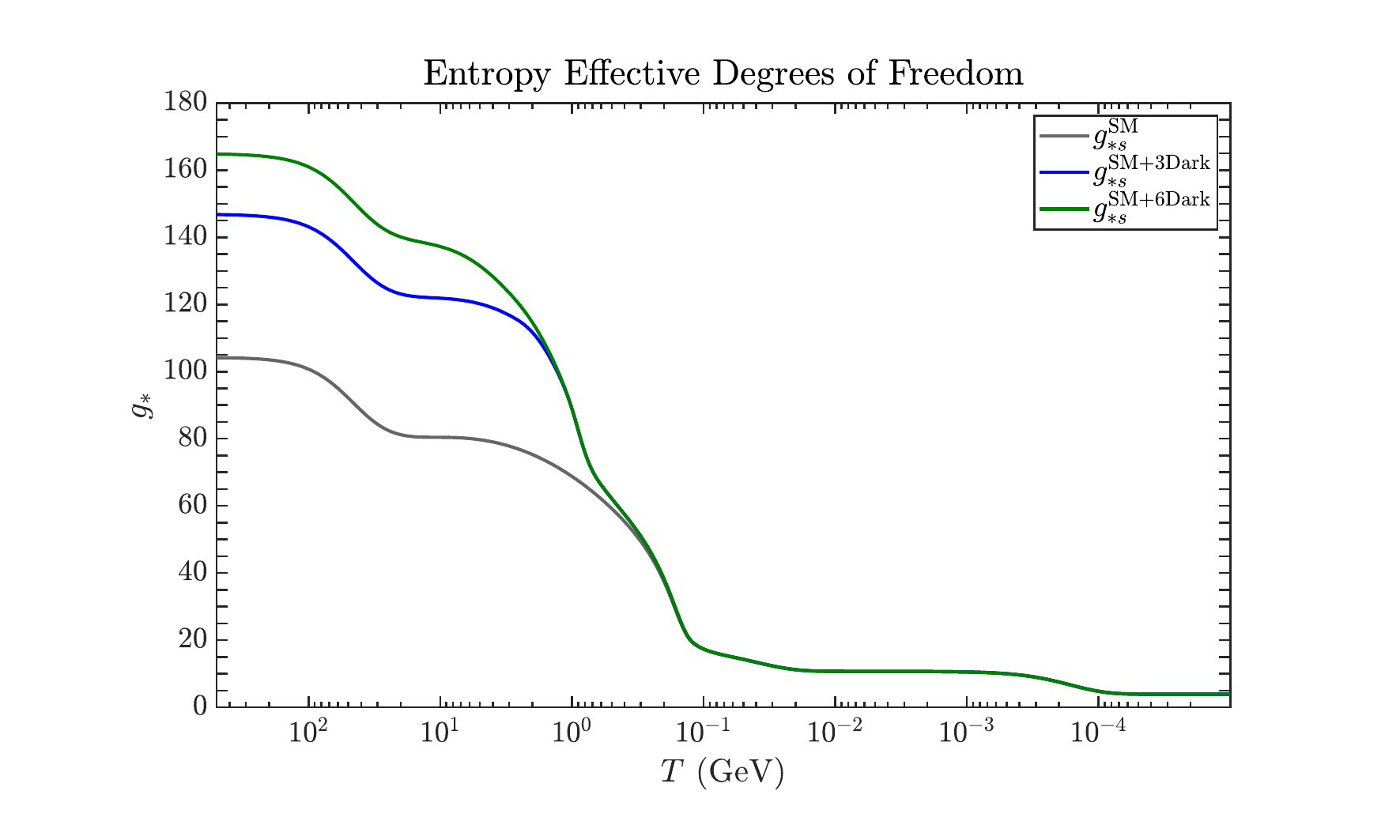}
        \caption{Total effective entropy degrees of freedom as functions of temperature for the pure SM (grey),
        the SM plus a minimal dark QCD sector with $(u^\prime,d^\prime,s^\prime)$ (blue),
        and the SM plus a dark QCD sector with all six dark quark flavors (dark green).
        The effective degrees of freedom of the eight dark gluons are included for both ${\rm SM+dQCD}_3$ and ${\rm SM+dQCD}_6$.}
		\label{fig: SdQCD}
\end{figure}

The total effective energy degrees of freedom, including the minimal dark QCD sector with three light flavors, is plotted in Fig.~\ref{fig: EdQCD}.
The effective entropy degrees of freedom are obtained analogously using Eq.~\eqref{eq:SD} and are shown in Fig.~\ref{fig: SdQCD}.

\paragraph*{SM plus a dark QCD sector with all six quark flavors  (${\rm SM+dQCD}_6$)}

We further extend the dark QCD sector to include six flavors of dark quarks $(u',d',s',c',b',t')$,
together with eight massless dark gluons.
The dark quark masses are again taken to be three times their SM counterparts,
and dark hadron masses are $5.5$ times the corresponding SM hadron masses.
Again, for simplicity, we do not include the dark Higgs in the analysis,
since its associated dynamics occur well before the crossover epoch
and are therefore not expected to significantly affect the resulting SIGW spectrum.

The procedure for computing the dark sector contribution to the effective relativistic degrees of freedom
follows that of the minimal dark QCD case discussed above.
For $T \lesssim 660~\mathrm{MeV}$ we compute the HRG contribution to the energy degrees of freedom.
Near the dark QCD crossover, we employ the SM lattice QCD fitting functions
for $\Delta(T)$ and $p(T)/T^4$ in the range $100~\mathrm{MeV}- 1~\mathrm{GeV}$ (for four quark flavors) from~\cite{Borsanyi:2016ksw},
and apply the $5.5$ temperature rescaling to obtain the dark quark and dark gluon contributions
for $660~\mathrm{MeV}\lesssim T\lesssim 5.5~\mathrm{GeV}$.

For the six-flavor dark QCD sector, we follow a procedure analogous to that used for the ${\rm SM+dQCD}_3$ case.
We use the $N_f=4$ perturbative result for the pressure.
Heavy quark mass effects are incorporated as in Section~\ref{sec:HQTh}. We again consider two switching temperatures, $T_s=2.75~{\rm GeV}$ and $T_s=3~{\rm GeV}$. For each value of $T_s$, we implement two matching prescriptions: (1) fixing the renormalization scale to $\mu=2\pi T$ and varying $q_c(N_f=4)$; and (2) fixing $q_c(N_f=4)=-3000$ and varying the renormalization scale over the range $\mu\in[\pi T,3\pi T]$. This procedure yields four curves for $\Delta(T)$. The corresponding effective degrees of freedom, $g_{\ast}(T)$ and $g_{\ast s}(T)$, are then obtained following the same steps as in the ${\rm SM+dQCD}_3$ case.

The total effective energy and entropy degrees of freedom including a dark QCD sector with all six quark flavors are also shown
in Figs.~\ref{fig: EdQCD} and~\ref{fig: SdQCD} respectively.

\section{Scalar induced gravitational waves}\label{Sec:SIGW}

In this work we use SIGWs as a probe of the thermal history around the QCD or the dark QCD crossovers. The basic idea is that an enhanced primordial curvature perturbation generates tensor perturbations at the second order after horizon reentry. Since the scalar transfer function is sensitive to the background equation of state and the speed of sound, a temporary change in these quantities around a crossover leaves an imprint on the resulting GW spectrum. In the following, we summarize the framework used in our numerical calculation. We work in the conformal Newtonian gauge and consider scalar perturbations at first order and tensor perturbations at second order~\cite{Zhou:2024doz}. The perturbed flat FLRW metric is written as
\begin{equation}
{\rm d}s^2
= a^2(\eta)\Big\{
-(1+2\phi){\rm d}\eta^2
+\big[(1-2\psi)\delta_{ij}
+\tfrac{1}{2}h_{ij}\big]{\rm d} x^i {\rm d} x^j
\Big\}\,,
\label{eq:SIGW_metric}
\end{equation}
where $\eta$ denotes the conformal time, with the relation between conformal time and temperature derived in Appendix~\ref{App:UTtoCT}; 
$\phi$ and $\psi$ are the first-order scalar perturbations, and $h_{ij}$ is the second-order tensor perturbation.
First-order vector and tensor perturbations are neglected.
The matter sector is modeled as an adiabatic perfect fluid,
\begin{equation}
T_{\mu\nu}=(\rho+p)\,u_\mu u_\nu+ w\,\rho\,g_{\mu\nu} \,,
\end{equation}
with equation of state $w=p/\rho$.
At the background level one obtains
\begin{equation}
3\mathcal{H}^2=\kappa\, a^2\rho \,,
\qquad
\mathcal{H}'=-\frac{1}{2}(1+3w)\mathcal{H}^2 \,,
\label{eq:background_FLRW}
\end{equation}
where $\mathcal{H}=a'/a$, $\kappa=8\pi G$, and a prime denotes differentiation with respect to conformal time. For a perfect fluid with negligible anisotropic stress, the two scalar potentials satisfy $\psi=\phi$. Then, the first-order scalar perturbation obeys
\begin{equation}
\phi_{\bf k}''(\eta)
+3\mathcal{H}(1+c_s^2)\phi_{\bf k}'(\eta)
+\left[
3(c_s^2-w)\mathcal{H}^2
+c_s^2 k^2
\right]\phi_{\bf k}(\eta)=0 \,.
\label{eq:scalar_eom}
\end{equation}
In the radiation-dominated limit, $w=c_s^2=1/3$, $c_s^2={\rm d}p/ {\rm d}\rho$ is adiabatic sound speed. Eq.~\eqref{eq:scalar_eom} reduces to
\begin{equation}
\phi_{\bf k}''+4\mathcal{H}\phi_{\bf k}'
+\frac{k^2}{3}\phi_{\bf k}=0 \,,
\end{equation}
with the solution
\begin{equation}
\phi_{\bf k}(\eta)=\psi_{\bf k}(\eta)
=\frac{2}{3}\zeta_{\bf k}\,T_\phi(|{\bf k}|\eta) \,,
\label{eq:phi_transfer_RD}
\end{equation}
where $\zeta_{\bf k}$ is the primordial curvature perturbation, $T_{\phi}(|\mathbf{k}|\eta)$ is the transfer function \cite{Kohri:2018awv, Zhou:2024doz}
\begin{eqnarray}\label{eq:T2}
		T_{\phi}(|\mathbf{k}|\eta)=\frac{9}{(|\mathbf{k}|\eta)^{2}}\Bigg[\frac{\sqrt{3}}{|\mathbf{k}|\eta} \sin \left(\frac{|\mathbf{k}| \eta}{\sqrt{3}}\right)-\cos \left(\frac{|\mathbf{k}| \eta}{\sqrt{3}}\right)\Bigg] \,.
\end{eqnarray}

The preceding results are valid for the evolution of first‑order scalar perturbations in the radiation‑dominated era. However, for the QCD or the dark QCD crossover, $w$ and $c_s^2$ vary with time. We therefore solve Eq.~\eqref{eq:scalar_eom} numerically, using the corresponding equation of state as input. We write the solution as
\begin{equation}
\phi_{\bf k}(\eta)=\frac{2}{3}\zeta_{\bf k}\,T_\phi(k,\eta)\,,
\end{equation}
with the early-time initial conditions
\begin{equation}
T_\phi( k,\eta_i)=1\,,
\qquad
T_\phi'( k,\eta_i)=0\,,
\end{equation}
where $\eta_i$ is chosen sufficiently before horizon reentry.

Once the explicit time‑dependent forms of the equation‑of‑state parameter $w$ and the sound‑speed parameter $c_s^2$ are specified, we can compute the first‑order transfer function under the initial conditions given above. After solving for the first‑order scalar perturbations, we can further evaluate the second‑order SIGWs, whose evolution equation can be written as
\begin{equation}
h_{ij}''+2\mathcal{H}h_{ij}'-\Delta h_{ij}
=
-4\Lambda_{ij}^{\ \ lm}\,\mathcal{S}_{lm}\,,
\label{eq:tensor_eom_position}
\end{equation}
where $\Lambda_{ij}^{\ \ lm}$ is the transverse-traceless projection operator. The source in Eq.~(\ref{eq:tensor_eom_position}) can be written as
\begin{align}
\mathcal{S}_{lm}
=&
\left[2-\frac{4}{3(1+w)}\right]
\partial_l\phi\,\partial_m\phi
-\frac{4}{3(1+w)\mathcal{H}}
\left(
\partial_l\phi'\,\partial_m\phi
+\partial_l\phi\,\partial_m\phi'
\right)\nonumber \\
&-\frac{4}{3(1+w)\mathcal{H}^2}
\partial_l\phi'\,\partial_m\phi'
+4\phi\,\partial_l\partial_m\phi \,.
\label{eq:SIGW_source}
\end{align}
We decompose the tensor perturbation into polarization modes,
\begin{equation}
h_{ij}({\bf x},\eta)
=
\sum_{\lambda=+,\times}
\int\frac{{\rm d}^3k}{(2\pi)^{3/2}}
{\rm e}^{i{\bf k}\cdot{\bf x}}
\varepsilon_{ij}^{\lambda}({\bf k})
h_{\bf k}^{\lambda}(\eta)\,,
\end{equation}
which leads to
\begin{equation}
h_{\bf k}^{\lambda\,\prime\prime}
+2\mathcal{H}h_{\bf k}^{\lambda\,\prime}
+k^2h_{\bf k}^{\lambda}
= 4\widetilde{S}^{\lambda}_{\bf k}(\eta)\,.
\label{eq:tensor_eom}
\end{equation}
Introducing $X_{\bf k}^{\lambda}=a h_{\bf k}^{\lambda}$, Eq.~\eqref{eq:tensor_eom} becomes
\begin{equation}
X_{\bf k}^{\lambda\,\prime\prime}
+\left(k^2-\frac{a''}{a}\right)X_{\bf k}^{\lambda}
=
4a(\eta)\widetilde{S}_{\bf k}^{\lambda}(\eta)\,.
\end{equation}
The solution can be expressed using the Green's function $G_k(\eta,\bar{\eta})$ defined by
\begin{equation}
G_k''(\eta,\bar{\eta})
+\left(k^2-\frac{a''}{a}\right)G_k(\eta,\bar{\eta})
=
\delta(\eta-\bar{\eta}) \,.
\end{equation}
Thus,
\begin{equation}
h_{\bf k}^{\lambda}(\eta)
=
\frac{4}{a(\eta)}
\int_{\eta_i}^{\eta}{\rm d}\bar{\eta}\,
G_k(\eta,\bar{\eta})a(\bar{\eta})
\widetilde{S}_{\bf k}^{\lambda}(\bar{\eta}) \,.
\label{eq:tensor_green_solution}
\end{equation}
The tensor power spectrum is defined by
\begin{equation}
\left\langle
h_{\bf k}^{\lambda}(\eta)
h_{{\bf k}'}^{\lambda'}(\eta)
\right\rangle
=
\frac{2\pi^2}{k^3}
\delta_{\lambda\lambda'}
\delta^{(3)}({\bf k}+{\bf k}')
\mathcal{P}_h(k,\eta)\ .
\end{equation}
Evaluating the two-point correlation function of the second-order SIGWs yields the corresponding tensor power spectrum,
which takes the form~\cite{Abe:2020sqb}
\begin{align}
\mathcal{P}_h(k,\eta)
=
\frac{64}{81a^2(\eta)}
\int_{0}^{\infty}{\rm d}v\int_{|1-v|}^{|1+v|}{\rm d}u\,
{I^2(k,v,u,\eta)}
\frac{\big[4u^2-(1-v^2+u^2)^2\big]^2}
{16u^2v^2}
\mathcal{P}_\zeta(vk)
\mathcal{P}_\zeta(uk)\,,
\label{eq:Ph_final}
\end{align}
where $u=k_1/k$, $v=k_2/k$, and $\mathcal{P}_\zeta(k)$ is the power spectrum of primordial curvature perturbation.
The kernel function in Eq.~(\ref{eq:Ph_final}) is given by
\begin{align}
I(k,k_1,k_2,\eta)
=&k^2\int_{\eta_i}^{\eta} {\rm d}\bar{\eta}\,
a(\bar{\eta})G_k(\eta,\bar{\eta})
\Bigg\{
2T_\phi(k_1,\bar{\eta})T_\phi(k_2,\bar{\eta})\nonumber \\
&+
\frac{4}{3[1+w(\bar{\eta})]}
\left[
T_\phi(k_1,\bar{\eta})
+\frac{T_\phi'(k_1,\bar{\eta})}{\mathcal{H}(\bar{\eta})}
\right]\left[
T_\phi(k_2,\bar{\eta})
+\frac{T_\phi'(k_2,\bar{\eta})}{\mathcal{H}(\bar{\eta})}
\right]\Bigg\}\,.
\label{eq:I_kernel}
\end{align}
Using the expression for the power spectrum of SIGWs, we can compute the corresponding energy density spectrum. The spectral GW energy density fraction at time $\eta$ is given by
\begin{equation}
	\Omega_{\rm GW}(k,\eta)
	\equiv \frac{\rho_{\mathrm{GW}}(\eta,k)}{\rho_{\mathrm{tot}}(\eta)}=\frac{\rho_{\mathrm{GW}}(\eta,k)}{3M_{\mathrm{Pl}}^{2}H^2[\eta(T)]}
	=\frac{1}{24}
	\left(\frac{k}{aH}\right)^2
	\overline{\mathcal{P}_h(k,\eta)} \,,
\label{eq:OmegaGW_eta}
\end{equation}
where $H(\eta)$ is the Hubble parameter evaluated at conformal time $\eta$, with the corresponding temperature determined through the relation $\eta=\eta(T)$. The overline denotes an oscillation average taken after the tensor mode has entered well inside the horizon.

The present-day spectrum is obtained by redshifting Eq.~\eqref{eq:OmegaGW_eta} from the production epoch to today. In our numerical analysis, the dark QCD effects enter through the time-dependent functions $w(\eta)$, $c_s^2(\eta)$, $a(\eta)$, and $\mathcal{H}(\eta)$.

For illustration and for later numerical scans, we also consider
a monochromatic primordial spectrum given in Eq.~\eqref{eq:mono_spectrum}.
The reference scale $k_\ast$ is chosen such that the corresponding scalar mode reenters the horizon near the dark QCD crossover. This choice maximizes the sensitivity of the SIGW spectrum to the variation of the sound speed during the crossover. For the monochromatic primordial spectrum, the power spectrum of SIGWs reduces to
\begin{equation}
\mathcal{P}_h(k,\eta)
= \frac{64}{81a^2(\eta)}
{I^2(k,k_\ast,k_\ast,\eta)}
\frac{\big(4k_\ast^2-k^2\big)^2}{16k^2k_\ast^2}
A_\zeta^2 \,,
\qquad
0<k<2k_\ast \,.
\label{eq:Ph_mono}
\end{equation}
The spectral shape and amplitude are then controlled by the primordial amplitude $A_\zeta$, the reentry scale $k_\ast$, and the crossover-induced evolution of $w$ and $c_s^2$. This provides the basis for using SIGWs as a probe of both the visible QCD crossover and possible dark QCD crossovers.

\section{Probing the dark QCD crossover}\label{Sec:result}

After inflationary reheating, the Universe enters a radiation-dominated epoch. As the Universe expands and cools, particle species gradually become non-relativistic once the temperature drops below their masses and subsequently decouple from the radiation bath. The radiation energy and entropy densities are conventionally expressed in terms of the effective relativistic degrees of freedom, $g_\ast(T)$ and $g_{\ast s}(T)$, as given in Eq.~\eqref{eq:EEden}.

The equation of state parameter $w$ and the adiabatic sound speed $c_s^2$ can then be written in terms of $g_\ast(T)$ and $g_{\ast s}(T)$ as
\begin{align}
    w(T)&=\frac{4\,g_{\ast s}(T)}{3\,g_\ast(T)} - 1 \ , \\
    c_s^2(T)&=\frac{4 \left[ g_{\ast s}'(T)\,T + 4\,g_{\ast s}(T) \right]}{3 \left[ g_\ast'(T)\,T + 4\,g_\ast(T) \right]} - 1 \ .
\end{align}
Here the effective degrees of freedom are defined with respect to the temperature of the radiation bath.
In particular, after neutrino decoupling, the neutrino temperature evolves independently of the photon temperature, and after $e^+e^-$ annihilation one has $T_\nu=(4/11)^{1/3}T_\gamma$. Since $g_\ast(T)$ and $g_{\ast s}(T)$ are conventionally defined with respect to the photon temperature, this temperature difference must be accounted for in the low-temperature regime.

As discussed in Section~\ref{Sec:gdQCD} and Appendix~\ref{App:SMg}, hadronic effects, especially near the QCD or the dark QCD crossover, dominate the nontrivial temperature dependence of $g_\ast(T)$ and $g_{\ast s}(T)$. These effects are determined from lattice QCD, supplemented by the hadron resonance gas description below the crossover and perturbative QCD above it. In the present analysis, the effects of both the SM QCD and dark QCD crossovers are incorporated through the corresponding evolution of $g_\ast(T)$ and $g_{\ast s}(T)$, and hence through $w(T)$ and $c_s^2(T)$.

It has been proposed that the hidden sector may contain a mirror copy of the SM~\cite{Lee:1956qn,Kobzarev:1966qya,Blinnikov:1982eh,Chacko:2005pe},
and that the visible and dark baryon asymmetries may share a common origin~\cite{An:2009vq,Farina:2015uea,Feng:2020urb}.
If the symmetric components annihilate before BBN, the remaining asymmetric dark baryons can constitute the dark matter.
In this framework, a dark QCD confinement scale about 5.5 times of the QCD confinement scale
naturally gives a dark baryon mass approximately $5.5$ times the SM proton mass. Since cogenesis motivates comparable baryon numbers in the two sectors, this mass ratio can account for the observed dark to baryon abundance ratio.

In the SM, the QCD crossover corresponds to the confinement of light quarks and gluons into hadronic degrees of freedom, leading to a rapid decrease in their effective relativistic contribution to the thermal bath. In the dark sector, the corresponding dark quarks and dark gluons undergo an analogous transition at the higher temperature. Since this occurs earlier, when the total radiation bath contains a larger number of relativistic degrees of freedom, the fractional change in the effective degrees of freedom induced by the dark QCD crossover is diluted relative to the SM QCD case. Consequently, for the benchmark considered here, the dark QCD crossover can leave a shifted imprint on the equation of state and the sound speed, and the resulting SIGW spectrum, while its effect is not expected to exceed the SM QCD contribution.

We focus on the temperature-dependent profiles of $g_\ast(T)$ and $g_{\ast s}(T)$ introduced in the previous sections.
In the SIGW analysis below, we compare the following thermal histories:
\begin{itemize}
\item the SM-only case, highlighting the effect of the SM QCD crossover on SIGW;
\item the SM plus a minimal dark QCD sector with three light flavors $({\rm SM+dQCD}_3)$.
\end{itemize}
The case ${\rm SM+dQCD}_6$ is also discussed in Section~\ref{Sec:gdQCD}. Its effect on $w(T)$ and $c_s^2(T)$ near the dark QCD crossover is close to that of the ${\rm SM+dQCD}_3$ benchmark, and the resulting SIGW spectrum does not show a significant qualitative difference. We therefore concentrate on the minimal three-flavor dark QCD benchmark ${\rm SM+dQCD}_3$ in the following.

\begin{figure}[htp]
	\centering
	\includegraphics[width=12cm]{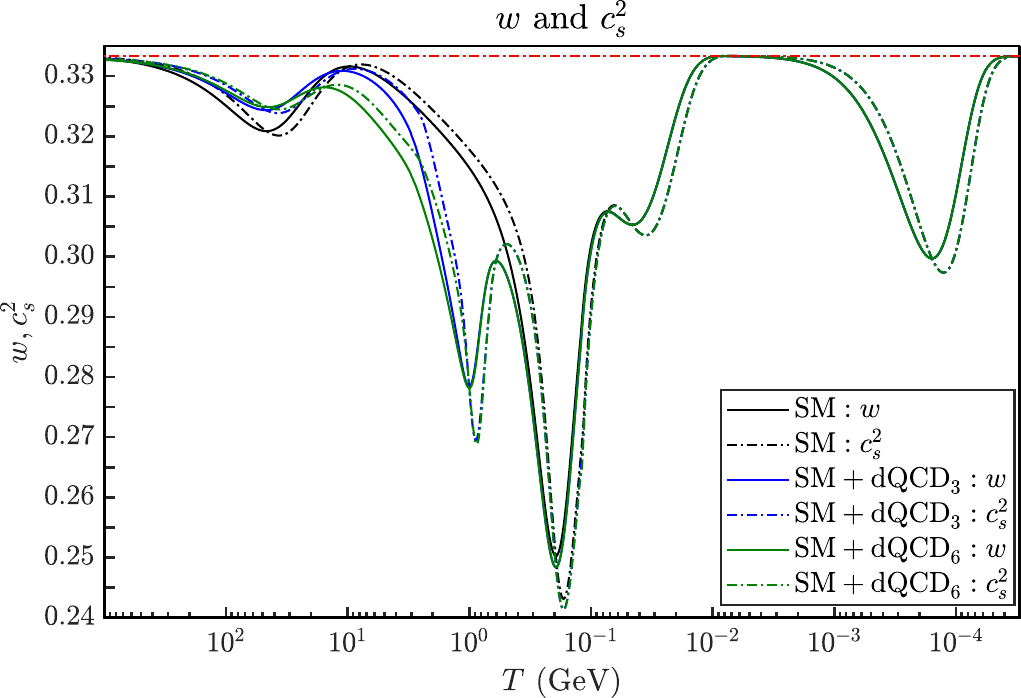}
	\caption{The equation of state parameter $w$ and sound speed $c_s^2$ as functions of temperature for the pure SM thermal bath (grey), the SM plus a minimal dark QCD sector with $(u^\prime,d^\prime,s^\prime)$ (blue), and the SM plus a dark QCD sector with all six dark quark flavors (dark green). The red dash-dotted line denotes the radiation-dominated limit, $w=c_s^2=1/3$.}
	\label{fig:cs_w}
\end{figure}

For each thermal history, we compute the resulting SIGW spectrum by solving the scalar transfer function through the QCD and dark QCD crossover epochs. The crossover effects enter through the time-dependent equation of state and sound speed, which modify the scalar perturbation dynamics and consequently the second-order tensor source.

\subsection{The SM QCD crossover effect}

SIGWs are sensitive to the detailed thermal history of the plasma. In particular, changes in the equation of state and sound speed caused by the QCD crossover modify the scalar transfer function and thus the resulting SIGW spectrum.
In this subsection, we isolate the effect of the SM QCD crossover by comparing the realistic SM thermal history with an idealized radiation-dominated background.

In a radiation-dominated  Universe with $w=c_s^2=1/3$, the SIGW energy density fraction for a monochromatic primordial curvature spectrum admits the analytic expression~\cite{Kohri:2018awv}
\begin{equation}
\begin{split}
\Omega_{\mathrm{GW}}(k)
=&\frac{3A_{\zeta }^2}{64}
\left[
\left(\frac{4-\tilde k^2}{4}\right)
(3\tilde k^2-2)\tilde k
\right]^2
\Bigg\{
\pi^2(3\tilde k^2-2)^2
\Theta(2\sqrt{3}-3\tilde k)\\
&\qquad\qquad\qquad
+\left[
4+(3\tilde k^2-2)
\log\left|1-\frac{4}{3\tilde k^2}\right|
\right]^2
\Bigg\}
\Theta(2-\tilde k)\,,
\end{split}
\label{eq:Omega_RD_analytic}
\end{equation}
where
\begin{equation}
\tilde k \equiv \frac{k}{k_*}=\frac{1}{u}=\frac{1}{v}.
\end{equation}
The analytic radiation-dominated spectrum has three characteristic features:
\begin{enumerate}
  \item \textbf{Singularity:} $\Omega_{\mathrm{GW}}$ contains a logarithmic singularity at
\begin{equation}
\tilde k_{\rm peak}
={2}/{\sqrt{3}}\,,
\end{equation}
which originates from resonant amplification: two scalar perturbations with wavenumber $k_*$ oscillate with frequency $2c_s k_*$, and resonance occurs when the tensor mode frequency satisfies
\begin{equation}
k
=2c_s k_*\,.
\end{equation}
For $c_s=1/\sqrt{3}$, this gives $\tilde k_{\rm peak}=2/\sqrt{3}$.
  \item \textbf{Zero point of the spectrum:} the radiation-dominated spectrum has a zero  at
\begin{equation}
k_{\rm zero}
=
\sqrt{2}\,c_s k_*\,
\qquad
\tilde k_{\rm zero}
=k_{\rm zero}/k_*=\sqrt{{2}/{3}}\,.
\end{equation}
  \item \textbf{Cut off:} the spectrum is strictly cut off for
\begin{equation}
\tilde k>2\,,
\end{equation}
since the integration region in Eq.~\eqref{eq:Ph_final} shrinks to zero measure,
as required by momentum conservation in the convolution of two scalar modes~\cite{Kohri:2018awv}.
\end{enumerate}
%
%

For constant $w$ and $c_s^2$, the kernel function can be expressed in terms of Ferrers functions and associated Legendre functions~\cite{Domenech:2021ztg}. In this case, the exact zero of $\Omega_{\mathrm{GW}}(k,\eta)$ occurs only for $w=1/3$. In the realistic SM thermal history, both $w$ and $c_s^2$ evolve with time and deviate from $1/3$, as shown in Fig.~\ref{fig:cs_w}. Then, the zero of the radiation-dominated spectrum and logarithmic singularity are modified: the zero becomes a finite local minimum, while the resonant peak is smoothed into a finite cusp. Moreover, the time evolution of $w(\eta)$ and $c_s^2(\eta)$ does not modify the cutoff of the energy‑density spectrum. Similar to the result in~\cite{Abe:2020sqb}, our calculation reveals a method for identifying the position of the modified trough in the energy density spectrum. Specifically, the trough is located approximately at $\sqrt{2}c_s(\eta_{\mathrm{cancel}})$, where $\eta_{\mathrm{cancel}}$ is the time determined by the condition $k_*\eta_{\mathrm{cancel}}\approx 4$.

\begin{figure}[t!]
\centering
\includegraphics[width=10cm]{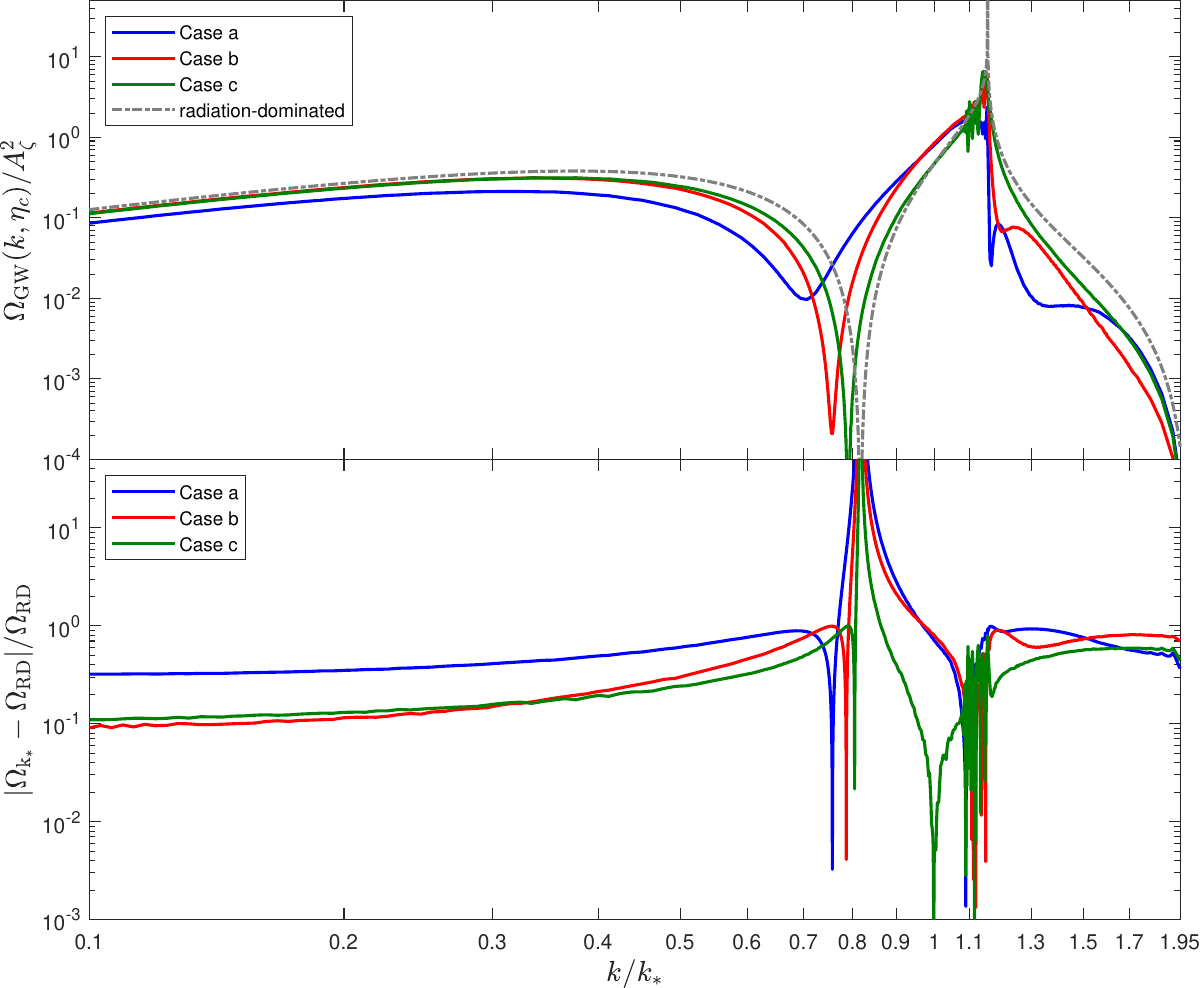}
\caption{Upper panel: SIGW energy density fraction
$\Omega_{\rm GW}(k,\eta_c)/A_\zeta^2$ for three representative choices of the monochromatic peak scale $k_*$, compared with the radiation-dominated result $w=c_s^2=1/3$ (black dashed line). The corresponding benchmark values are listed in Table~\ref{tab:SM_QCD_kstar}. All spectra are evaluated at $k_*\eta_c=400$. Lower panel: Absolute relative difference $|\Omega_{\rm model}-\Omega_{\rm RD}|/\Omega_{\rm RD}$ between each SM thermal-history result and the radiation-dominated case. }
\label{fig:spectrum1}
\end{figure}

To better understand how the SM QCD crossover affects the energy density spectrum of SIGWs, we consider a monochromatic primordial curvature perturbation and vary its peak scale $k_*$, with representative choices of $k_*$ summarized in Table~\ref{tab:SM_QCD_kstar}.
\textbf{Case a} probes the QCD crossover region, where $c_s^2$ reaches its minimum. \textbf{Case b} corresponds to a later time, when the sound speed has already begun to recover. Furthermore, \textbf{Case c} corresponds to a much later time, well after the QCD crossover, and is therefore closest to the radiation-dominated limit. We then calculate the corresponding energy density spectra $\Omega_{\rm GW}(k,\eta_c)/A_\zeta^2$ and compare them with the radiation-dominated result. All spectra are evaluated at $k_*\eta_c=400$, which is sufficiently late for the GW energy density fraction to reach its asymptotic value after horizon reentry. The results of the energy density spectra are shown in Fig.~\ref{fig:spectrum1}. The corrections to the energy density spectrum are mainly concentrated in the trough region. Specifically, the zero of $\Omega_{\mathrm{GW}}(k)$ is replaced by a trough, whose shifted position corresponds to $\sqrt{2}c_s(\eta_{\mathrm{cancel}})$. Furthermore, the singularity in the energy density spectrum becomes a finite peak.





\begin{table}[htp]
\centering
\renewcommand{\arraystretch}{1.25}
\begin{tabular}{c c c p{1.6cm}}
\hline
\textbf{Case}
& $k_*$ [$\mathrm{Mpc}^{-1}$]
& $T(\eta_{\rm cancel})$ [GeV]
& $c_s^2(\eta_{\rm cancel})$\\
\hline

\textbf{a}
& $9.035\times 10^6$
& $0.1689$
& $0.2431$ \\

\textbf{b}
& $4.770\times 10^6$
& $0.1006$
& $0.2913$\\

\textbf{c}
& $9.773\times 10^5$
& $0.02146$
& $0.3146$ \\
\hline
\end{tabular}
\caption{Representative choices of the monochromatic peak scale $k_*$ used to study the SM QCD crossover effect on the energy density spectrum. Here, $\sqrt{2}c_s(\eta_{\mathrm{cancel}})$ corresponds to the location of the trough in the energy density spectrum.}
\label{tab:SM_QCD_kstar}
\end{table}


\subsection{SM QCD versus dark QCD}
\label{subsec:SM_vs_dQCD}

We now investigate the impact of the dark QCD crossover on the energy density spectrum of SIGW in the ${\rm SM+dQCD}_3$ case. In this paper, we have also computed $\Omega_{\rm GW}(k,\eta_c)/A_\zeta^2$ for the ${\rm SM+dQCD}_6$. However, as shown in Fig.~\ref{fig:cs_w}, the corresponding $w(T)$ and $c_s^2(T)$ profiles are very close to those of the ${\rm SM+dQCD}_3$ below $1.2~{\rm GeV}$. Since the temperature range most relevant for our analysis lies largely below this scale, the resulting SIGW spectra are nearly identical. We therefore do not display the ${\rm SM+dQCD}_6$ results separately.

\begin{figure}[!htp]
    \centering
    \includegraphics[width=10cm]{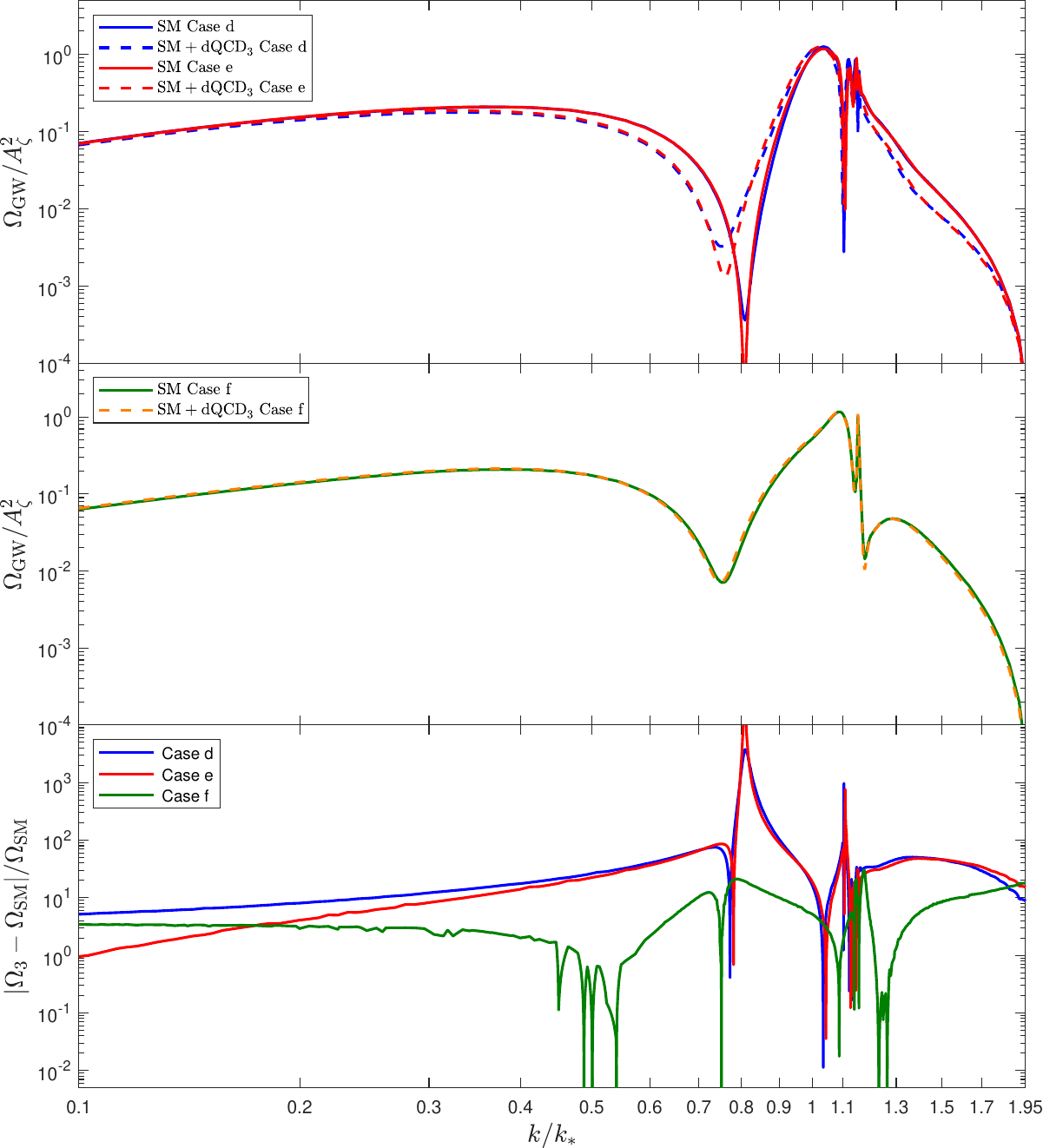}
    \caption{Upper panel and Middle panel: SIGW energy density spectra $\Omega_{\rm GW}(k,\eta_c)/A_\zeta^2$ for the SM thermal history (solid lines) and the ${\rm SM+dQCD}_3$ thermal history (dashed lines), for three representative choices of the monochromatic peak scale $k_*$. All spectra are evaluated at $k_*\eta_c=400$. Lower panel: Absolute relative difference
    $|\Omega_{{\rm SM+dQCD}_3}-\Omega_{\rm SM}|/\Omega_{\rm SM}$ between the ${\rm SM+dQCD}_3$ and SM results for each $k_*$. }
    \label{fig:spectrum2}
\end{figure}

We consider a monochromatic primordial spectrum and choose three representative values of  $k_\ast$ probing different thermal epochs, as summarized in Table~\ref{tab:DARK_QCD_kstar}.
\textbf{Cases d} and \textbf{e} both probe the dark QCD crossover region. \textbf{Case d} is chosen such that $c_s^2$ is near its minimum, whereas \textbf{Case e} adopts a value of $k_*$ that is $5.5$ times that used in~\cite{Abe:2020sqb}, where the reference value was chosen to probe the SM QCD crossover. This rescaling is motivated by the benchmark relation introduced in Eq.~(\ref{eq:QCDdQCD}), and correspondingly shifts the probed epoch to the dark QCD crossover region.
\textbf{Case f} corresponds to a lower temperature of about $0.3~{\rm GeV}$. As shown in Fig.~\ref{fig:cs_w}, the dark QCD corrections are already small relative to the SM contributions at this temperature. Consequently, the corresponding modification of the energy-density spectrum in Fig.~\ref{fig:spectrum2} is also negligible.






\begin{table}[htp]
\centering
\renewcommand{\arraystretch}{1.25}
\begin{tabular}{c c c p{1.6cm}}
\hline
\textbf{Case}
& $k_*$ [$\mathrm{Mpc}^{-1}$]
& $T(\eta_{\rm cancel})$ [GeV]
& $c_s^2(\eta_{\rm cancel})$\\
\hline

\textbf{d}
& $5.753\times 10^7$
& $0.8871$
& $0.2695$ \\

\textbf{e}
& $5.225\times 10^7$
& $0.8173$
& $0.2725$\\

\textbf{f}
& $1.830\times 10^7$
& $0.3081$
& $0.2909$ \\

\hline
\end{tabular}
\caption{Representative choices of the monochromatic peak scale $k_*$ used to study the SM QCD crossover effect on the energy density spectrum. The $T$-$\eta$ relation differs between the $\mathrm{SM+dQCD}_{3}$ and SM cases. The temperatures listed in the table correspond to the $\eta$ values of the $\mathrm{SM+dQCD}_{3}$ case.}
\label{tab:DARK_QCD_kstar}
\end{table}

The results of the energy density spectra are shown in Fig.~\ref{fig:spectrum2}. For \textbf{Cases d} and~\textbf{e}, the finite dip values of $\Omega_{\rm GW}(k,\eta_c)/A_\zeta^2$ in the ${\rm SM+dQCD}_3$ case are significantly larger than those in the corresponding SM thermal history. This behavior is consistent with the trend observed in Fig.~\ref{fig:spectrum1}: when $c_s^2(\eta_{\rm cancel})$ lies near the bottom of a sound speed dip, the zero of the radiation-dominated spectrum is lifted to a larger finite value. Furthermore, \textbf{Case f} corresponds to a lower temperature, around $0.3~{\rm GeV}$, which lies below the dark QCD crossover region. At the cancellation time, determined by $k_*\eta_{\rm cancel}\sim 4$, the values of $w$ and $c_s^2$ in the SM and ${\rm SM+dQCD}3$ scenarios are already close to each other. Consequently, the corresponding spectra $\Omega_{\rm GW}(k,\eta_c)/A_\zeta^2$ are also close.

\section{Conclusion}\label{Sec:Con}

In this work we studied SIGWs as probes of a dark QCD crossover. SIGWs are sensitive to the scalar transfer function, and hence to the equation of state and sound speed of the cosmic plasma. A hidden confining sector can therefore leave an indirect but characteristic imprint on the stochastic gravitational wave spectrum, even if the corresponding dark hadrons are difficult to access experimentally.

Motivated by twin Higgs and asymmetric twin baryon dark matter scenarios, we considered a dark QCD sector with an enhanced confinement scale approximately $5.5$ times the SM QCD scale. This benchmark naturally places the dark QCD crossover at a temperature above the SM QCD crossover. We derived the effective energy and entropy degrees of freedom for two representative scenarios: a minimal dark QCD sector with three light dark quark flavors and a dark QCD sector containing all six dark quark flavors. From the resulting $g_{\ast}(T)$ and $g_{\ast s}(T)$, we determined the corresponding equation of state parameter $w(T)$ and sound speed $c_s^2(T)$.

We then computed the SIGW spectrum by solving the scalar transfer function with the full time dependence of $w(\eta)$ and $c_s^2(\eta)$ and using the resulting scalar perturbations as the second-order source for tensor modes. Compared with the radiation-dominated background, the realistic SM thermal history modifies the analytic SIGW spectrum: the exact zero of the radiation-dominated result is lifted to a finite local minimum, while the logarithmic resonant singularity is smoothed into a finite cusp. By varying the peak scale $k_*$ of a monochromatic primordial curvature spectrum, we showed that the size of this modification depends on when the scalar mode enters the horizon relative to the QCD-induced dip in $w$ and $c_s^2$.

We further compared the SM thermal history with the SM supplemented by
a minimal dark QCD sector with three light flavors, ${\rm SM+dQCD}_3$. When the selected scalar peak scale probes the dark QCD crossover region, the corresponding SIGW spectrum shows a clear deviation from the pure SM result. In particular, the finite dip value of $\Omega_{\rm GW}(k,\eta_c)/A_\zeta^2$ is enhanced
when the cancellation time lies near the bottom of the dark QCD induced dip in the sound speed.
By contrast, when the mode enters the horizon below the dark QCD crossover region, the SM and ${\rm SM+dQCD}_3$ spectra remain close to each other, since the relevant values of $w$ and $c_s^2$ are similar at the cancellation time. We also checked that the six-flavor dark QCD benchmark ${\rm SM+dQCD}_6$ gives a SIGW spectrum qualitatively similar to the ${\rm SM+dQCD}_3$ benchmark in the temperature range most relevant for the dark QCD crossover.

The characteristic location of the dark QCD imprint is set by the horizon-entry scale associated with the dark QCD crossover. For the benchmark considered here in Eq.~\eqref{eq:QCDdQCD}, the corresponding feature is shifted to a larger wavenumber, or equivalently to a higher present-day frequency, relative to the SM QCD feature. A useful way to present this comparison is to use a common reference scale, for example $k_{\rm QCD}$ or $f_{\rm QCD}$, and thus the pure SM QCD feature appears near $k/k_{\rm QCD}\sim 1$, while the dark QCD feature appears near $k/k_{\rm QCD}\sim 5.5$, up to corrections from the changing effective degrees of freedom and entropy evolution.

Our results suggest that SIGWs provide a complementary cosmological probe of hidden confining sectors. Since horizon reentry around the QCD scale corresponds to frequencies relevant for pulsar timing arrays (PTA), future PTA observations may be sensitive not only to the SM QCD imprint but also to frequency-shifted features associated with a dark QCD crossover.
A statistically robust search would require fitting SIGW templates incorporating the SM and dark QCD thermal histories to PTA spectral data and comparing them with smooth astrophysical backgrounds and templates based on the SM thermal history alone.
If a dark QCD sector exists near the benchmark scale motivated by asymmetric twin baryon dark matter, its crossover could manifest as a shifted spectral distortion in the SIGW background, providing a gravitational wave probe of otherwise hidden strong dynamics.
The analysis developed here can also be extended to other well-motivated frameworks containing additional dark confining sectors with different numbers of quark flavors, quark masses, and couplings.\\

\noindent\textbf{Acknowledgments:}

WZF is supported in part by the National Natural Science Foundation of China under Grant No. 11935009,
and Tianjin University Self-Innovation Fund Extreme Basic Research Project Grant No. 2025XJ21-0007.

\appendix

\section{Effective degrees of freedom of the SM}\label{App:SMg}

In this appendix, we review the calculation of the energy and entropy effective degrees of freedom for the SM thermal bath. Since the present work focuses on the QCD crossover and the mirror dark QCD crossover, with the latter motivated to occur at a scale approximately $5.5$ times of the SM QCD scale, we restrict our discussion to the temperature regime below the electroweak crossover that is most relevant for these phenomena.

\subsection{Effective degrees of freedom in the lepton-photon era}

For temperatures around the $e^\pm$ annihilation epoch,
several subleading effects are incorporated to determine
the precise parametrization of the effective degrees of freedom:
\begin{itemize}
  \item Corrections from non-instantaneous neutrino decoupling.
  \item The entropy release from $e^\pm$ annihilation and the resulting evolution of the photon-neutrino temperature ratio.
  \item Finite-mass contributions from species such as $e^\pm$, $\mu^\pm$.
  \item Finite-temperature QED corrections to the electromagnetic plasma.
  \item Contributions from light hadrons.
\end{itemize}
For $T\lesssim 10~\mathrm{MeV}$,
the effective energy and entropy degrees of freedom are written as~\cite{Saikawa:2018rcs}
\begin{align}
	g_{\ast}(T)
	&=
	2a_{r1}
	+\frac{21}{4}\left(\frac{4}{11}\right)^{\frac{4}{3}}
	a_{r2}\,
	\mathcal{S}^{\frac{4}{3}}\!\left(\frac{m_e}{T}\right)
	+\frac{\rho^0_e(T)+\rho_{\rm QED}(T)+\rho_{\mu}^0(T)+\rho_{\rm hadrons}(T)}{{\pi^2T^4}/{30}}\,,
	\label{eq:gastee}
	\\
	g_{\ast s}(T)
	&=
	2a_{s1}
	+\frac{21}{11}\,a_{s2}\,\mathcal{S}\!\left(\frac{m_e}{T}\right)
	+\frac{s^0_e(T)+s_{\rm QED}(T)+s_{\mu}^0(T)+s_{\rm hadrons}(T)}{{2\pi^2T^3}/{45}}\,.
	\label{eq:gastSee}
\end{align}
In the above equations, the coefficients $a_{r1}, a_{r2}, a_{s1}, a_{s2}$
parametrize deviations from instantaneous decoupling.
The values $a_{r1}\simeq 1.014$ and $a_{r2}\simeq 0.9947$ are fixed by requiring
the first two terms on the right-hand side of Eq.~(\ref{eq:gastee}) to
approach $g_{\ast,\slashed{\nu}}=3.383$
in the limit $T\ll m_e$~\cite{Dolgov:1998sf,Mangano:2001iu,Mangano:2005cc,deSalas:2016ztq,Saikawa:2018rcs} ,
and $2+6\cdot\frac{7}{8}$ in the limit $T\gg m_e$.
Similarly, $a_{s1}\simeq 1.017$ and $a_{s2}\simeq 0.9935$ are fixed by requiring
the first two terms on the right-hand side of Eq.~(\ref{eq:gastSee}) to
approach $g_{\ast x,\slashed{\nu}}=3.931$ for $T\ll m_e$~\cite{Dolgov:1998sf,Mangano:2001iu,Mangano:2005cc,deSalas:2016ztq,Saikawa:2018rcs},
and $2+6\cdot\frac{7}{8}$ for $T\gg m_e$.

The function $\mathcal{S}$ is introduced to track the annihilation transition, defined by
\begin{equation}
		\mathcal{S}(y) \equiv 1+ \frac{15y^5}{2\pi^4}I_{F2}(y)\,,
\end{equation}
where the fermion thermal integrals entering this expression are
\begin{equation}
		I_{F1}(y) \equiv \int^{\infty}_1 \frac{(x^2-1)^{\frac{3}{2}}}{{\rm e}^{yx}+1} \,{\rm d}x\,,\qquad
		I_{F2}(y) \equiv \int^{\infty}_1 \frac{{\rm e}^{yx}(x^2-1)^{\frac{3}{2}}}{({\rm e}^{yx}+1)^2}x \,{\rm d}x\,. \label{eq:IF}
\end{equation}

The contribution of a single particle species $i$ to the energy and entropy densities in the ideal gas approximation is given by
\begin{align}
		\rho_i^0(T) &= \Delta_{F,i}^0(T) T^4 + {3p_{F,i}^0(T)} \label{rho_fermion_free}\,,\\
		s_i^0(T) &= \Delta_{F,i}^0(T) T^3 + {4p^0_{F,i}(T)}/{T} \label{s_fermion_free}\,,
\end{align}
where the pressure and the trace anomaly derived from the pressure are given by
\begin{align}
		p_{F,i}^0(T) &= \frac{g_i}{6\pi^2}m_i^4 I_{F1}\left(\frac{m_i}{T}\right)\,,\label{p_fermion_free} \\
		\Delta_{F,i}^0(T) &= \frac{g_i}{6\pi^2}\frac{m_i^4}{T^4}\left[\frac{m_i}{T}I_{F2}\left(\frac{m_i}{T}\right)-4I_{F1}\left(\frac{m_i}{T}\right)\right]\,. \label{delta_fermion_free}
\end{align}

Finite-temperature QED corrections are incorporated via
\begin{align}
		\rho_{\rm QED}(T) &= \Delta_{\rm QED}(T) T^4 + {3p_{\rm QED}(T)}\,,\\
		s_{\rm QED}(T) &= \Delta_{\rm QED}(T) T^3 + {4p_{\rm QED}(T)}/{T}\,,
\end{align}
where $p_{\rm QED}(T)$ denotes the pressure, and $\Delta_{\rm QED}(T)$ is the associated trace anomaly,
\begin{align}
		p_{\rm QED}(T) &= -\int^{\infty}_0\frac{{\rm d}k}{2\pi^2}\left[\frac{k^2}{E_k}\frac{\delta m_e^2(T)}{{\rm e}^{E_k/T}+1}+\frac{k}{2}\frac{\delta m_{\gamma}^2(T)}{{\rm e}^{k/T}-1}\right]\,,\\
		\Delta_{\rm QED}(T) &= T\frac{{\rm d}}{{\rm d}T}\left[\frac{p_{\rm QED}(T)}{T^4}\right]\,.\label{delta_QED}
\end{align}
Here the thermal mass corrections are
\begin{align}
		\delta m_e^2(T) &= \frac{2\pi\alpha T^2}{3} + \frac{4\alpha}{\pi}\int^{\infty}_0
{\rm d}k\frac{k^2}{E_k}\frac{1}{{\rm e}^{E_k/T}+1}\,,\label{finite_T_correction_m_e} \\
		\delta m_{\gamma}^2(T) &= \frac{8\alpha}{\pi}\int^{\infty}_0 {\rm d}k \frac{k^2}{E_k}\frac{1}{{\rm e}^{E_k/T}+1}\,.
\end{align}

In Eqs.~(\ref{eq:gastee}) and~(\ref{eq:gastSee}),
the contributions of muons and hadrons ($\rho_{\rm hadrons}, s_{\rm hadrons}$)
are included to accomplish a smooth connection to the calculations at higher temperatures.
The hadronic contributions $\rho_{\rm hadrons}$ and $s_{\rm hadrons}$ are given by
Eqs.~(\ref{eq:p_hadrons}) and~(\ref{eq:Del_hadrons}).


\subsection{Passing the QCD crossover}

The contribution of strongly interacting species to the effective number of relativistic degrees of freedom
can be organized into three temperature regimes:
(1) the hadronic phase well below the crossover, $T \ll 10^{2}\,{\rm MeV}$;
(2) the non-perturbative regime in the vicinity of the crossover, $T \sim 10^{2}\,{\rm MeV}$~\cite{Saikawa:2018rcs};
and (3) the perturbative quark--gluon phase well above the QCD crossover, $T \gg 10^{2}\,{\rm MeV}$.
These regimes are then matched smoothly to provide a unified description across the full temperature range.

For strongly interacting species in a thermal medium, estimating the energy density by a naive counting of particle degrees of freedom is no longer reliable, since the plasma cannot be accurately approximated as an ideal gas. Instead, the effective degrees of freedom should be extracted from the thermodynamics of the interacting plasma. In principle, all thermodynamic quantities can be obtained from the pressure, provided that $p(T)$ is differentiable over the temperature range of interest. Since the pressure may be difficult to determine directly and unambiguously, one commonly computes the trace anomaly $I(T)$, reconstructs the pressure from it, and then derives the temperature-dependent effective degrees of freedom, $g_{*}(T)$ and $g_{*s}(T)$.

\paragraph{Trace anomaly}

For a reversible process,
\begin{equation}
{\rm d}U=T{\rm d}S-p{\rm d}V\,,
\end{equation}
and thus
\begin{equation}
{\rm d}S=\frac{1}{T}\left({\rm d}U+p{\rm d}V\right)\,,
\end{equation}
writing $U=\rho V$ and $S= s V$, one obtains
\begin{align}
{\rm d}(sV) & =\frac{1}{T}\left[{\rm d}(\rho V)+p{\rm d}V\right]\,.
\end{align}
Comparing the ${\rm d}V$ and ${\rm d}/{\rm d}T$ terms one has
\begin{align}
s(T) & =\frac{1}{T}\left[\rho(T)+p(T)\right]\,,\\
\rho(T) & =T\frac{{\rm d}p}{{\rm d}T}-p(T)\,.
\end{align}
Thus, the energy and entropy densities can be obtained from the pressure via standard thermodynamic relations.
In practice, this requires knowledge of the pressure and its temperature derivative,
while properly accounting for interactions among SM particles.
For this purpose, it is convenient to introduce the dimensionless trace anomaly
\begin{equation}
\Delta(T)\equiv\frac{\rho(T)-3p(T)}{T^{4}}=T\frac{{\rm d}}{{\rm d}T}\left[\frac{p(T)}{T^{4}}\right]\,, \label{eq:DimlTA}
\end{equation}
which relates the trace of the stress-energy $T_{\mu}^{\mu}=\rho-3p$.
Once $p$ and $\Delta$ are obtained, the energy and entropy densities can be derived as
\begin{align}
\rho(T) & =\Delta(T) T^{4} +{3p(T)}\,, \label{eq:ED}\\
s(T) & = \Delta(T) T^{3} + {4p(T)}/{T}\,, \label{eq:SD}
\end{align}
and $g_{\ast}(T)$ and $g_{\ast s}(T)$ can be thus obtained from
\begin{equation}
\rho(T)=\frac{\pi^{2}}{30}g_{\ast}(T)T^{4}\,,\qquad s(T)=\frac{2\pi^{2}}{45}g_{\ast s}(T)T^{3}\,. \label{eq:EEden}
\end{equation}
Accordingly, one reconstructs $p(T)$ by integrating the trace anomaly $\Delta(T)$ with appropriate boundary conditions.
Starting from the definition of the trace anomaly in Eq.~(\ref{eq:DimlTA}),
integrating both sides gives
\begin{equation}
	\frac{p(T)}{T^{4}}=\frac{p(T_{0})}{T_{0}^{4}}+\int_{T_{0}}^{T}\frac{{\rm d}T^{\prime}}{T^{\prime}}\,\Delta(T^{\prime})\,. \label{eq:Deltop}
\end{equation}
Here $T_{0}$ is a low reference temperature at which $p(T_{0})$ is known
(often taking $p(T_{0})\to 0$, so that the integration constant is negligible).
The energy density then follows as Eq.~(\ref{eq:ED}).

\paragraph{After the QCD crossover}

After the QCD crossover, the thermodynamics of the hadronic plasma is commonly described within the hadron resonance gas model (HRG),
in which the medium is approximated as a non-interacting gas of hadrons and resonances~\cite{Hagedorn:1965st,Dashen:1969ep,Venugopalan:1992hy}.
This description is in good agreement with lattice QCD results~\cite{Borsanyi:2013bia,Bellwied:2015lba}.
In the HRG model, the dominant attractive interactions in the hadronic phase are effectively accounted for by treating resonances as additional particle species.
Accordingly, rather than introducing explicit interaction terms, one extends the ideal gas sum to include the resonance spectrum.
The pressure and energy density then follow from the standard phase space integrals, from which the trace anomaly can be computed.
In addition to the light hadronic states, baryons and mesons with masses below $2.5~{\rm GeV}$ are included in the following sum
\begin{align}
	p_{\rm hadrons}(T) &= \sum_i p_i^0(T)\,,\label{eq:p_HRG}\\
	\Delta_{\rm hadrons}(T) &= \sum_i \Delta_i^0(T)\,, \label{eq:delta_HRG}
\end{align}
where the pressure of each species is
\begin{align}
	p_i^0(T) =
	\left\{
	\begin{array}{ll}
		p_{B,i}^0(T) = {\displaystyle\frac{g_i}{6\pi^2}m_i^4 I_{B1}\left(\frac{m_i}{T}\right)} & \text{for bosons}\,, \\ [2.0ex]
		p_{F,i}^0(T) = {\displaystyle\frac{g_i}{6\pi^2}m_i^4 I_{F1}\left(\frac{m_i}{T}\right)} & \text{for fermions}\,, \\
	\end{array}
	\right.
\end{align}
and the corresponding trace anomaly contributions are
\begin{align}
	\Delta_i^0(T) = \left\{
	\begin{array}{ll}
		\Delta_{B,i}^0(T) = {\displaystyle \frac{g_i}{6\pi^2}\frac{m_i^4}{T^4}\left[\frac{m_i}{T}I_{B2}\left(\frac{m_i}{T}\right)-4I_{B1}\left(\frac{m_i}{T}\right)\right]} & \text{for bosons}\,, \\ [2.0ex]
		\Delta_{F,i}^0(T) = {\displaystyle \frac{g_i}{6\pi^2}\frac{m_i^4}{T^4}\left[\frac{m_i}{T}I_{F2}\left(\frac{m_i}{T}\right)-4I_{F1}\left(\frac{m_i}{T}\right)\right]} & \text{for fermions}\,, \\
	\end{array}
	\right.
\end{align}
In the above equations, boson thermal integrals are defined as
\begin{equation}
	I_{B1}(y) \equiv \int^{\infty}_1 \frac{(x^2-1)^{3/2}}{{\rm e}^{yx}-1}\,{\rm d}x\,,\qquad
	I_{B2}(y) \equiv \int^{\infty}_1 \frac{{\rm e}^{yx}(x^2-1)^{3/2}}{({\rm e}^{yx}-1)^2}x\,{\rm d}x\,, \label{eq:IB}
\end{equation}
and fermion thermal integrals are given in Eq.~(\ref{eq:IF}).

Hence, the energy and entropy densities can be computed as
\begin{align}
\rho_{\rm hadrons}(T) & =\Delta_{\rm hadrons}(T) T^{4} +{3\,p_{\,\rm hadrons}(T)}\,, \label{eq:p_hadrons}\\
s_{\rm hadrons}(T) & = \Delta_{\rm hadrons}(T) T^{3} +{4\,p_{\,\rm hadrons}(T)}/{T}\,, \label{eq:Del_hadrons}
\end{align}
which also enter in Eqs.~(\ref{eq:gastee}) and~(\ref{eq:gastSee}).

\paragraph{Near the QCD crossover}

The hadron resonance gas description ceases to be reliable in the vicinity of the QCD crossover.
In this regime, quark--gluon interactions are strongly coupled and perturbation theory breaks down.
One therefore relies on lattice QCD to determine the thermodynamic properties of the quark--gluon plasma,
from which the pressure and trace anomaly can be extracted and thus the effective number of relativistic degrees of freedom can be obtained.

In lattice QCD simulations, one often starts by including only the light quark flavors.
Early results for $2+1$ flavor ($u,d,s$) QCD were reported in~\cite{Borsanyi:2010cj,Borsanyi:2013bia} and later confirmed by~\cite{HotQCD:2014kol}.
Subsequently, the charm quark was incorporated, extending the analysis to $2+1+1$ flavor QCD~\cite{Borsanyi:2016ksw},
thereby enabling an estimate of the equation of state up to temperatures of order $1~{\rm GeV}$.
In this work, we therefore adopt the $2+1+1$ flavor lattice QCD results.

\paragraph{Before the QCD crossover}

Before the crossover, the thermodynamics can be described using finite-temperature perturbative QCD (pQCD).
The high-temperature pressure of QCD with (effectively) massless quarks has been computed
within an effective field theory framework~\cite{Ginsparg:1980ef,Braaten:1995jr,Appelquist:1981vg},
which provides a systematic way to combine perturbative contributions with non-perturbative results in a consistent manner.
The coefficients in the expansion in the strong coupling $g_s$ have been determined at successive orders by~\cite{Shuryak:1977ut,Chin:1978gj,Kapusta:1979fh,Toimela:1982hv,Arnold:1994ps,Arnold:1994eb,Zhai:1995ac,Braaten:1995jr,Kajantie:2002wa,Ellis:2023oxs}.
In this work, we adopt the result truncated at $\mathcal{O}\!\left(g_s^{6}\ln\frac{1}{g_s}\right)$~\cite{Kajantie:2002wa},
\begin{equation}
	p_{\rm QCD}(T) = \frac{8\pi^2}{45}T^4 \left[p_0 + p_2\frac{\alpha_s}{\pi}+p_3\left(\frac{\alpha_s}{\pi}\right)^{\frac{3}{2}}
	+ p_4\left(\frac{\alpha_s}{\pi}\right)^2 + p_5\left(\frac{\alpha_s}{\pi}\right)^{\frac{5}{2}} + p_6\left(\frac{\alpha_s}{\pi}\right)^3 \right]\,,
\label{eq:p_pQCD}
\end{equation}
where the coefficients $p_i$ depend on the number of (effectively) massless quark flavors $N_f$,
the renormalization scale $\mu$ in the modified $\overline{\rm MS}$ scheme, and the temperature, and are given by
\begin{align}
	p_0 &= 1 + \frac{21}{32}N_f\,,\quad	p_2 = -\frac{15}{4}\left(1+ \frac{5}{12}N_f\right)\,, \quad
	p_3 = 30\left(1+\frac{1}{6}N_f\right)^{\frac{3}{2}}\,, \\
	p_4 &= 237.2 + 15.96 N_f - 0.4150 N_f^2 + \frac{135}{2}\left(1+\frac{1}{6}N_f\right)\ln\left[\frac{\alpha_s}{\pi}
\left(1+\frac{1}{6}N_f\right)\right] \nonumber\\
	&\quad -\frac{165}{8}\left(1+\frac{5}{12}N_f\right)\left(1-\frac{2}{33}N_f\right)\ln\left(\frac{\mu}{2\pi T}\right)\,, \\
	p_5 &= \left(1+\frac{1}{6}N_f\right)^{\frac{1}{2}} \left[-799.1-21.96N_f -1.926N_f^2
+ \frac{495}{2}\left(1+\frac{1}{6}N_f\right)\left(1-\frac{2}{33}N_f\right)\ln\left(\frac{\mu}{2\pi T}\right)\right]\,, \\
	p_6 &= \left[-659.2 - 65.89N_f - 7.653 N_f^2 + \frac{1485}{2}\left(1+\frac{1}{6}N_f\right)\left(1-\frac{2}{33}N_f\right)\ln\left(\frac{\mu}{2\pi T}\right)\right]\nonumber\\
	&\quad \times \ln\left[\frac{\alpha_s}{\pi}\left(1+\frac{1}{6}N_f\right)\right] -475.6\ln\left(\frac{\alpha_s}{\pi}\right) - \frac{1815}{16}\left(1+\frac{5}{12}N_f\right)\left(1-\frac{2}{33}N_f\right)^2\ln^2\left(\frac{\mu}{2\pi T}\right) \nonumber\\
	&\quad + \left(2932.9 + 42.83N_f - 16.48N_f^2 + 0.2767N_f^3\right)\ln\left(\frac{\mu}{2\pi T}\right) + q_c(N_f)\,. \label{eq:p_QCD_order_g6}
\end{align}
The coefficient $p_6$ involves a function $q_c(N_f)$,
reflecting the infrared sensitivity of finite-temperature field theory (the Linde's problem)~\cite{Linde:1980ts,Gross:1980br}.
In practice, one treats $q_c$ as a phenomenological parameter that
effectively captures the net impact of higher-order terms,
and fixes it by matching to lattice QCD results.

The contribution to the trace anomaly can be thus evaluated as
\begin{equation}
\Delta_{\rm QCD}(T)=T\frac{{\rm d}}{{\rm d}T}\left[\frac{p_{\rm QCD}(T)}{T^{4}}\right]\,,
\end{equation}
using which the energy and entropy densities can be obtained.

\paragraph{Heavy quark thresholds}\label{sec:HQTh}

The perturbative expression for the QCD pressure is strictly applicable only when the relevant quark flavors can be treated as (approximately) massless.
This assumption becomes questionable near heavy-quark thresholds.
For charm with $m_c\simeq 1.3\,\mathrm{GeV}$, the massless approximation can induce sizable deviations when $T\sim m_c$.
Finite-mass corrections to the QCD pressure have been computed up to $\mathcal{O}(g_s^2)$ in~\cite{Laine:2006cp},
where higher-order effects in the mass dependence were found to be subdominant compared to higher-order perturbative corrections to the pressure itself.
Motivated by this observation, it was argued that heavy-quark threshold effects can be captured adequately by a tree-level correction factor~\cite{Borsanyi:2016ksw}.
In this work we follow this prescription and incorporate quark mass effects accordingly.

Concretely, the pressure of the $(udsc)$ system
including the effect of the charm quark mass as
\begin{equation}
	p_{\rm QCD}^{(udsc)}(T)
	=
	\frac{p_{\rm QCD,SB}(T,3)+p_{F,c}^0(T)}{p_{\rm QCD,SB}(T,4)}
	\left. p_{\rm QCD}(T)\right|_{N_f=4}\,,\label{eq:Pressure4q}
\end{equation}
where
\begin{equation}
	p_{\rm QCD,SB}(T,N_f)=\frac{\pi^2}{90}T^4\left(16+\frac{21}{2}N_f\right)
\end{equation}
is the Stefan--Boltzmann pressure of the $N_f$--flavor theory,
$p_{F,c}^{0}(T)$ denotes the ideal gas contribution from a free charm quark,
and $p_{\rm QCD}(T)|_{N_f=4}$  corresponds to Eq.~(\ref{eq:p_pQCD}) evaluated at $N_f = 4$.

Including the bottom quark ($m_b\simeq 4.2~\mathrm{GeV}$) can be treated analogously,
\begin{equation}
	p_{\rm QCD}^{(udscb)}(T)
	=
	\frac{p_{\rm QCD,SB}(T,4)+p_{F,b}^0(T)}{p_{\rm QCD,SB}(T,4)}
	p_{\rm QCD}^{(udsc)}(T)\,,\label{eq:Pressure5q}
\end{equation}
with $p_{F,b}^{0}(T)$ the free bottom-quark contribution.
Hence, the contribution to the trace anomaly can be estimated as
\begin{equation}
\Delta_{\rm QCD}^{(udscb)}(T)=T\frac{{\rm d}}{{\rm d}T}\left[\frac{p_{\rm QCD}^{(udscb)}(T)}{T^{4}}\right]\,,
\label{eq: DimlTA}
\end{equation}

For the top quark, being the heaviest ($m_t\simeq 172~\mathrm{GeV}$), the regime $T\gg m_t$ implies that all quarks are effectively massless (and one must also bear in mind that electroweak symmetry is restored at sufficiently high temperatures). In that limit, one may use the pQCD result with $N_f=6$, while ensuring a smooth transition by interpolating the trace anomaly,
\begin{equation}
	\Delta_{\rm QCD}^{(udscbt)}(T)
	=
	c_t(T) \Delta_{\rm QCD}(T)\big|_{N_f=6}
	+
	\left[1-c_t(T)\right]\Delta_{\rm QCD}^{(udscb)}(T)\,,
	\label{eq:delta_QCD6}
\end{equation}
where
\begin{equation}
	c_t(T)=\frac{p_{F,t}^0(T)\big|_{m_t=m_t(v_T)}}{\frac{7\pi^2}{60}T^4}\,.
	\label{c_topmass}
\end{equation}

\subsection{Complete equation of state and associated uncertainties}

\paragraph*{Hadronic sector contribution}

In the intermediate regime, $T=\mathcal{O}(0.1$--$10)\,\mathrm{GeV}$,
the perturbative expansion of the QCD pressure is poorly convergent,
and thus one reconstructs it from the trace anomaly via Eq.~(\ref{eq:Deltop}).
We evaluate $p(T_\ast)/T_\ast^4$ at $T_\ast=10\,\mathrm{MeV}$ and integrate upward,
and independently estimate $p(T_\ast)/T_\ast^4$ at $T_\ast=10^{17}\,\mathrm{GeV}$ from the perturbative SM result and integrate downward. 
The integrand is decomposed as $\Delta=\Delta_{\rm leptons}+\Delta_{\rm strong}+\Delta_{\rm electroweak}$,
where QED corrections are included only for $T\le120\,\mathrm{MeV}$.

The dominant uncertainty originates from the QCD contribution $\Delta_{\rm strong}(T)$.
For the low-temperature integration, the trace anomaly in the strongly interacting sector is written as~\cite{Saikawa:2018rcs}
\begin{equation}
	\Delta_{\rm strong}(T)=
	\left\{
	\begin{array}{ll}
		\Delta_{\rm hadrons}(T) & \text{for}\quad T \le 120~\mathrm{MeV}\,, \\[0.5ex]
		\Delta_{\rm lattice}^{(2+1+1)}(T) & \text{for}\quad 120~\mathrm{MeV} < T \le T_s\,, \\[0.5ex]
		\Delta_{\rm QCD}^{(udscbt)}(T) & \text{for}\quad T > T_s\,,
	\end{array}
	\right.
	\label{delta_strong}
\end{equation}
where $\Delta_{\rm hadrons}$ is obtained from the HRG description in Eq.~(\ref{eq:delta_HRG}),
$\Delta_{\rm lattice}^{(2+1+1)}$ from lattice QCD,
and $\Delta_{\rm QCD}^{(udscbt)}$ from the pQCD result including heavy quark threshold effects, given in
Eq.~\eqref{eq:delta_QCD6},
with $T_s=500\,\mathrm{MeV}$ or $1\,\mathrm{GeV}$.

The lattice--pQCD matching uncertainty is then assessed using two prescriptions at each switching temperature $T_s$:
(1) fixing $\mu=2\pi T$ and varying $q_c(N_f=4)$, and (2) fixing $q_c(N_f=4)=-3000$ while varying $\mu$.
The associated error band is then taken as the envelope of the reconstructed results.
The low- and high-temperature integrations overlap around $T\sim\mathcal{O}(10)\,\mathrm{GeV}$.
Within this window, they are stitched together using an interpolating ansatz for $p(T)$,
yielding a smooth, unified equation of state across the full temperature range~\cite{Saikawa:2018rcs}.

\paragraph*{High-temperature corrections to the effective degrees of freedom}

As pointed out in~\cite{Saikawa:2018rcs}, even at temperatures well above the electroweak crossover
the effective relativistic degrees of freedom do not exactly approach the ideal gas value
$g_{\ast}=g_{\ast s}=106.75$.
This can be understood from perturbative corrections to the pressure.
Writing the weak coupling expansion as
\begin{equation}
  \frac{p(T,g)}{T^{4}}=\hat p_{0}+\hat p_{2}\,g^{2}+\cdots,
\end{equation}
and noting that the first non-vanishing contribution to the trace anomaly appears only at $\mathcal{O}(g^{4})$,
the leading deviation of $g_{\ast i}$ from its tree-level value can be estimated
directly from the $\mathcal{O}(g^{2})$ correction to the pressure.
One then estimates
\begin{equation}
  \delta g_{\ast i}
  \;\equiv\;
  g_{\ast i}-g_{\ast i,\mathrm{tree}}
  \;=\;
  \frac{90}{\pi^{2}}\,\hat p_{2}\,g^{2}+\cdots,
  \qquad (i=\rho,s),
\end{equation}
where $g_{\ast i,\mathrm{tree}}=106.75$ in the SM.
In QCD, $\hat p_{2}=(2/45)\,p_{2}$ with
\begin{equation}
  p_{2}=-\frac{15}{4}\left(1+\frac{5}{12}N_{f}\right)<0\,,
\end{equation}
so that $\delta g_{\ast i}<0$.
For $N_{f}=6$, one has $\hat p_{2}=-7/12$; taking $g_{s}^{2}\simeq 0.29$ at $T\sim 10^{15}\,\mathrm{GeV}$
yields the estimate $\delta g_{\ast i}\simeq -1.5$~\cite{Kajantie:2002wa,Davoudiasl:2004gf,Saikawa:2018rcs}.

\paragraph*{Leptonic sector contribution from 10~MeV to the electroweak crossover}

In this temperature range, the leptonic contribution to the trace anomaly is modeled as that of an ideal gas
of massive charged leptons, supplemented by a finite-temperature QED correction at low temperatures,
\begin{equation}
\Delta_{\rm leptons}(T)=
\begin{cases}
\Delta^{0}_{F,e}(T)+\Delta^{0}_{F,\mu}(T)+\Delta^{0}_{F,\tau}(T)+\Delta_{\rm QED}(T)\,,
& T\le 120~{\rm MeV},\\[0.6ex]
\Delta^{0}_{F,e}(T)+\Delta^{0}_{F,\mu}(T)+\Delta^{0}_{F,\tau}(T)\,,
& T>120~{\rm MeV},
\end{cases}
\end{equation}
with the corresponding pressure
\begin{equation}
p_{\rm leptons}(T)=p^{0}_{F,e}(T)+p^{0}_{F,\mu}(T)+p^{0}_{F,\tau}(T)\,.
\end{equation}
Accordingly, we include $e$, $\mu$, and $\tau$ as free fermions with their physical masses,
retain $\Delta_{\rm QED}(T)$ only for $T\le 120~{\rm MeV}$, and neglect it at higher temperatures.

While photons and (effectively) massless neutrinos satisfy $\Delta=0$ in the ideal gas limit,
they still contribute to the total pressure and thus affect the effective degrees of freedom.

For $T\gtrsim 10~{\rm MeV}$, the full trace anomaly is obtained by summing the contributions from the relevant sectors.
In the intermediate regime one may schematically write
\begin{equation}
\Delta_{\rm intermediate}(T)=\Delta_{\rm leptons}(T)+\Delta_{\rm strong}(T)+\Delta_{\rm electroweak,low}(T)\,,
\end{equation}
where $\Delta_{\rm strong}(T)$ denotes the QCD contribution (HRG/lattice/pQCD, with appropriate matching),
and $\Delta_{\rm electroweak,low}(T)$ accounts for the electroweak sector below the crossover.

\paragraph*{Final fitting functions}

We now provide fitting functions that reproduce $g_{*}$ and $g_{*s}$ for
$T \le 10^{16}~\mathrm{GeV}$.
For the range $120~\mathrm{MeV} \le T \le 10^{16}~\mathrm{GeV}$,
the fits are well approximated by~\cite{Saikawa:2018rcs}
\begin{align}
	g_{*}(T) &\simeq \frac{\sum_{i=0}^{11} a_i\, t^i}{\sum_{i=0}^{11} b_i\, t^i}\,,\label{gsr_fitting_function}\\
	\frac{g_{*}(T)}{g_{*s}(T)} &\simeq 1 + \frac{\sum_{i=0}^{11} c_i\, t^i}{\sum_{i=0}^{11} d_i\, t^i}\,,\label{gsr_to_gss_fitting_function}
\end{align}
where $t \equiv \ln (T~[\mathrm{GeV}])$. The coefficients $a_i$, $b_i$, $c_i$, and $d_i$
are listed in Table~\ref{tab:fit}.

\begin{table}[h]
	\caption{Coefficients for the fitting functions for $120\,\mathrm{MeV}\leq T \leq 10^{16}\,\mathrm{GeV}$. \label{tab:fit}}
	\centering \begin{tabular}{|c|rrrr|}
		\hline
		$i$&
		\multicolumn{1}{c}{$a_i$}&
		\multicolumn{1}{c}{$b_i$}&
		\multicolumn{1}{c}{$c_i$}&
		\multicolumn{1}{c|}{$d_i$}\\
		\hline
		\hline
		0&1&1.43382$\times10^{-2}$&1&7.07388$\times10^{1}$\\
		1&1.11724$\times10^{0}$&1.37559$\times10^{-2}$&6.07869$\times10^{-1}$&9.18011$\times10^{1}$\\
		2&3.12672$\times10^{-1}$&2.92108$\times10^{-3}$&$-$1.54485$\times10^{-1}$&3.31892$\times10^{1}$\\
		3&$-$4.68049$\times10^{-2}$&$-$5.38533$\times10^{-4}$&$-$2.24034$\times10^{-1}$&$-$1.39779$\times10^{0}$\\
		4&$-$2.65004$\times10^{-2}$&$-$1.62496$\times10^{-4}$&$-$2.82147$\times10^{-2}$&$-$1.52558$\times10^{0}$\\
		5&$-$1.19760$\times10^{-3}$&$-$2.87906$\times10^{-5}$&2.90620$\times10^{-2}$&$-$1.97857$\times10^{-2}$\\
		6&1.82812$\times10^{-4}$&$-$3.84278$\times10^{-6}$&6.86778$\times10^{-3}$&$-$1.60146$\times10^{-1}$\\
		7&1.36436$\times10^{-4}$&2.78776$\times10^{-6}$&$-$1.00005$\times10^{-3}$&8.22615$\times10^{-5}$\\
		8&8.55051$\times10^{-5}$&7.40342$\times10^{-7}$&$-$1.69104$\times10^{-4}$&2.02651$\times10^{-2}$\\
		9&1.22840$\times10^{-5}$&1.17210$\times10^{-7}$&1.06301$\times10^{-5}$&$-$1.82134$\times10^{-5}$\\
		10&3.82259$\times10^{-7}$&3.72499$\times10^{-9}$&1.69528$\times10^{-6}$&7.83943$\times10^{-5}$\\
		11&$-$6.87035$\times10^{-9}$&$-$6.74107$\times10^{-11}$&$-$9.33311$\times10^{-8}$&7.13518$\times10^{-5}$\\
		\hline
	\end{tabular}
\end{table}

For $T<120\,\mathrm{MeV}$, the fitting functions are written as
\begin{align}
	&g_{*}(T) \simeq 2.030 +1.353 \mathcal{S}_{\rm fit}^{\frac{4}{3}}\left(\frac{m_e}{T}\right) + 3.495
	f_\rho\left(\frac{m_e}{T}\right)+ 3.446 f_\rho\left(\frac{m_{\mu}}{T}\right) + 1.05b_\rho\left(\frac{m_{\pi^0}}{T}\right)\nonumber \\
	&
	+2.08 b_\rho\left(\frac{m_{\pi^{\pm}}}{T}\right)
	+4.165 b_\rho\left(\frac{m_{1}}{T}\right)
	+30.55 b_\rho\left(\frac{m_{2}}{T}\right)
	+89.4 b_\rho\left(\frac{m_{3}}{T}\right)
	+ 8209 b_\rho\left(\frac{m_{4}}{T}\right)
	,\label{gsr_fitting_function_low}\\
	&g_{*s}(T) \simeq 2.008 +1.923 \mathcal{S}_{\rm fit}\left(\frac{m_e}{T}\right) + 3.442
	f_s\left(\frac{m_e}{T}\right)+ 3.468 f_s\left(\frac{m_{\mu}}{T}\right) + 1.034b_s\left(\frac{m_{\pi^0}}{T}\right) \nonumber \\
	&
	+2.068 b_s\left(\frac{m_{\pi^{\pm}}}{T}\right)
	+4.16 b_s\left(\frac{m_{1}}{T}\right)
	+30.55 b_s\left(\frac{m_{2}}{T}\right)
	+90 b_s\left(\frac{m_{3}}{T}\right)
	+6209 b_s\left(\frac{m_{4}}{T}\right),\label{gss_fitting_function_low}
\end{align}
where $m_{e}=511\times10^{-6}$~GeV, $m_{\mu}=0.1056$~GeV, $m_{\pi^{0}}=0.135$~GeV, $m_{\pi^{\pm}}=0.140$~GeV, $m_{1}=0.5$~GeV, $m_{2}=0.77$~GeV, $m_{3}=1.2$~GeV, and $m_{4}=2$~GeV.
Functions used in Eqs.~\eqref{gsr_fitting_function_low} and~\eqref{gss_fitting_function_low} are given by
\begin{align}
	f_{\rho}(x) &= \exp(- 1.04855 x)(1 + 1.03757 x + 0.508630 x^2 + 0.0893988x^3 )\,,\\
	b_{\rho}(x) &= \exp(-1.03149 x)(1 +1.03317 x + 0.398264 x^2 + 0.0648056x^3 )\,,\\
	f_{s}(x) &=\exp(-1.04190 x)(1 +1.03400x + 0.456426 x^2 + 0.0595248x^3 )\,,\\
	b_{s}(x) &=\exp(-1.03365 x)(1 +1.03397 x + 0.342548 x^2 + 0.0506182x^3 )\,,\\
	\mathcal{S}_{\rm fit}(x) &= 1 + \tfrac{7}{4} \exp(- 1.0419 x)(1 + 1.034 x + 0.456426x^2 + 0.0595249x^3 )\,.
\end{align}

\section{Cosmic time to conformal time conversion}
\label{App:UTtoCT}

The Hubble parameter and conformal time are defined as
\begin{equation}
H(t)=\frac{\dot a(t)}{a(t)}\,,
\qquad
{\rm d}\eta=\frac{{\rm d}t}{a(t)}\,,
\end{equation}
with the present-day normalization $a_0=1$. In a flat
$\Lambda$CDM universe,
\begin{equation}
H(z)
=
H_0
\sqrt{
\Omega_\Lambda^0
+
\Omega_m^0(1+z)^3
+
\Omega_r^0(1+z)^4
}\,.
\end{equation}
For $z\gtrsim10^4$, the matter and dark energy contributions are negligible.
However, when evolving far back into the radiation era,
changes in the relativistic particle content must be included.

Conservation of comoving entropy gives
\begin{equation}
s_r(T)a^3(T)=s_{r0} a_0^3 = s_{r0}\,,
\end{equation}
and thus
\begin{equation}
a(T)
=
\frac{T_{\gamma0}}{T}
\left[
\frac{g_{*s,0}}{g_{*s}(T)}
\right]^{1/3}\,,
\label{eq:a_of_T_corrected}
\end{equation}
where \(T_{\gamma0}\) is the present CMB temperature.

The measured present radiation density is
\begin{equation}
\rho_{r0}
=
\frac{3H_0^2}{8\pi G}\Omega_r^0\,,
\end{equation}
and we define
\begin{equation}
g_{*,0}
\equiv
\frac{30\rho_{r0}}{\pi^2T_{\gamma0}^4}\,.
\end{equation}
Using
\begin{equation}
\frac{\rho_r(T)}{\rho_{r0}}
=
\frac{g_{*}(T)}{g_{*,0}}
\left(
\frac{T}{T_{\gamma0}}
\right)^4\,,
\end{equation}
the Friedmann equation during radiation domination gives
\begin{equation}
H(T)
=
H_0\sqrt{\Omega_r^0}
\left[
\frac{g_{*}(T)}{g_{*,0}}
\right]^{1/2}
\left(
\frac{T}{T_{\gamma0}}
\right)^2\,.
\label{eq:H_of_T_corrected}
\end{equation}

The evolution of the temperature follows from entropy conservation
\begin{equation}
g_{*s}(T)T^3a^3(T)
=
{\rm constant}\,,
\end{equation}
and differentiating this relation with respect to time yields
\begin{equation}
\frac{{\rm d}t}{{\rm d}T}
=
-
\frac{1}{H(T)T}
\left[
1+
\frac{T}{3g_{*s}(T)}
\frac{{\rm d}g_{*s}(T)}{{\rm d}T}
\right]\,.
\label{eq:dtdT_gs}
\end{equation}
Choosing $t(T\rightarrow\infty)=0$,
the cosmic time corresponding to temperature $T$ is therefore
\begin{equation}
t(T)
=
\frac{T_{\gamma0}^2}
{H_0\sqrt{\Omega_r^0}}
\int_T^\infty
\left[
\frac{g_{*,0}}
{g_{*}(T^{\prime})}
\right]^{1/2}
\left[
1+
\frac{T^{\prime}}
{3g_{*s}(T^{\prime})}
\frac{{\rm d}g_{*s}(T^{\prime})}
{{\rm d}T^{\prime}}
\right]
\frac{{\rm d}T^{\prime}}{{T^{\prime }}^3}\,.
\label{eq:t_of_T_gstar_corrected}
\end{equation}

Using  Eqs.~\eqref{eq:a_of_T_corrected}, \eqref{eq:H_of_T_corrected}, and \eqref{eq:dtdT_gs},
as well as the conformal time definition
\begin{equation}
\eta(T)
=
\int_0^{t(T)}
\frac{{\rm d}t'}{a(t')}\,,
\end{equation}
and changing the integration variable from $t'$ to $T^{\prime}$,
one obtains
\begin{equation}
\eta(T)
=
\frac{T_{\gamma0}}
{H_0\sqrt{\Omega_r^0}}
\int_T^\infty
\left[
\frac{g_{*s}(T^{\prime})}
{g_{*s,0}}
\right]^{1/3}
\left[
\frac{g_{*,0}}
{g_{*}(T^{\prime})}
\right]^{1/2}
\left[
1+
\frac{T^{\prime}}
{3g_{*s}(T^{\prime})}
\frac{{\rm d}g_{*s}(T^{\prime})}
{{\rm d}T^{\prime}}
\right]
\frac{{\rm d}T^{\prime}}{{T^{\prime}}^2}\,.
\label{eq:eta_of_T_corrected}
\end{equation}

If \(g_{*}\) and \(g_{*s}\) are approximately constant over the
relevant temperature interval, these expressions reduce to
\begin{equation}
t(T)
=
\frac{1}
{2H_0\sqrt{\Omega_r^0}}
\left(
\frac{T_{\gamma0}}{T}
\right)^2
\left[
\frac{g_{*,0}}{g_{*}(T)}
\right]^{1/2}\,,
\label{eq:t_constant_g}
\end{equation}
and
\begin{equation}
\eta(T)
=
\frac{1}
{H_0\sqrt{\Omega_r^0}}
\frac{T_{\gamma0}}{T}
\left[
\frac{g_{\ast s}(T)}{g_{*\ast s,0}}
\right]^{1/3}
\left[
\frac{g_{\ast,0}}{g_{\ast}(T)}
\right]^{1/2}\,.
\label{eq:eta_constant_g}
\end{equation}

The familiar ideal radiation relations
\begin{equation}
t
=
\frac{a^2}{2H_0\sqrt{\Omega_r^0}},
\qquad
\eta
=
\frac{a}{H_0\sqrt{\Omega_r^0}},
\qquad
t
=
\frac12H_0\sqrt{\Omega_r^0}\,\eta^2\,,
\end{equation}
are recovered when the radiation composition remains fixed.
Neglecting matter and dark energy for $z\gtrsim10^4$ remains valid.
The correction required for an accurate temperature conversion is the
temperature dependence of $g_{*\rho}$ and $g_{*s}$.



\begin{thebibliography}{99}
\bibitem{Saikawa:2018rcs}
K.~Saikawa and S.~Shirai,
JCAP \textbf{05}, 035 (2018)
doi:10.1088/1475-7516/2018/05/035
[arXiv:1803.01038 [hep-ph]].

\bibitem{Ananda:2006af}
K.~N.~Ananda, C.~Clarkson and D.~Wands,
Phys. Rev. D \textbf{75}, 123518 (2007)
doi:10.1103/PhysRevD.75.123518
[arXiv:gr-qc/0612013 [gr-qc]].

\bibitem{Baumann:2007zm}
D.~Baumann, P.~J.~Steinhardt, K.~Takahashi and K.~Ichiki,
Phys. Rev. D \textbf{76}, 084019 (2007)
doi:10.1103/PhysRevD.76.084019
[arXiv:hep-th/0703290 [hep-th]].

\bibitem{Domenech:2019quo}
G.~Dom{\`e}nech,
Int. J. Mod. Phys. D \textbf{29}, no.03, 2050028 (2020)
doi:10.1142/S0218271820500285
[arXiv:1912.05583 [gr-qc]].

\bibitem{Abe:2020sqb}
K.~T.~Abe, Y.~Tada and I.~Ueda,
JCAP \textbf{06}, 048 (2021)
doi:10.1088/1475-7516/2021/06/048
[arXiv:2010.06193 [astro-ph.CO]].

\bibitem{Domenech:2021ztg}
G.~Dom{\`e}nech,
Universe \textbf{7}, no.11, 398 (2021)
doi:10.3390/universe7110398
[arXiv:2109.01398 [gr-qc]].

\bibitem{Yuan:2021qgz}
C.~Yuan and Q.~G.~Huang,
iScience \textbf{24}, 102860 (2021)
doi:10.1016/j.isci.2021.102860
[arXiv:2103.04739 [astro-ph.GA]].

\bibitem{Balaji:2023ehk}
S.~Balaji, G.~Dom{\`e}nech and G.~Franciolini,
JCAP \textbf{10}, 041 (2023)
doi:10.1088/1475-7516/2023/10/041
[arXiv:2307.08552 [gr-qc]].

\bibitem{Zhu:2023gmx}
Q.~H.~Zhu, Z.~C.~Zhao, S.~Wang and X.~Zhang,
Chin. Phys. C \textbf{48}, no.12, 125105 (2024)
doi:10.1088/1674-1137/ad79d5
[arXiv:2307.13574 [astro-ph.CO]].

\bibitem{Liu:2023pau}
L.~Liu, Z.~C.~Chen and Q.~G.~Huang,
JCAP \textbf{11}, 071 (2023)
doi:10.1088/1475-7516/2023/11/071
[arXiv:2307.14911 [astro-ph.CO]].

\bibitem{Harigaya:2023pmw}
K.~Harigaya, K.~Inomata and T.~Terada,
Phys. Rev. D \textbf{108}, no.12, 123538 (2023)
doi:10.1103/PhysRevD.108.123538
[arXiv:2309.00228 [astro-ph.CO]].

\bibitem{Domenech:2024rks}
G.~Dom{\`e}nech, S.~Pi, A.~Wang and J.~Wang,
JCAP \textbf{08}, 054 (2024)
doi:10.1088/1475-7516/2024/08/054
[arXiv:2402.18965 [astro-ph.CO]].

\bibitem{An:2009vq}
H.~An, S.~L.~Chen, R.~N.~Mohapatra and Y.~Zhang,
JHEP \textbf{03}, 124 (2010)
doi:10.1007/JHEP03(2010)124
[arXiv:0911.4463 [hep-ph]].

\bibitem{Farina:2015uea}
M.~Farina,
JCAP \textbf{11}, 017 (2015)
doi:10.1088/1475-7516/2015/11/017
[arXiv:1506.03520 [hep-ph]].

\bibitem{Feng:2020urb}
W.~Z.~Feng and J.~H.~Yu,
Commun. Theor. Phys. \textbf{75}, no.4, 045201 (2023)
doi:10.1088/1572-9494/acbb5b
[arXiv:2005.06471 [hep-ph]].

\bibitem{Chacko:2005pe}
Z.~Chacko, H.~S.~Goh and R.~Harnik,
Phys. Rev. Lett. \textbf{96}, 231802 (2006)
doi:10.1103/PhysRevLett.96.231802
[arXiv:hep-ph/0506256 [hep-ph]].

\bibitem{Burdman:2006tz}
G.~Burdman, Z.~Chacko, H.~S.~Goh and R.~Harnik,
JHEP \textbf{02}, 009 (2007)
doi:10.1088/1126-6708/2007/02/009
[arXiv:hep-ph/0609152 [hep-ph]].

\bibitem{Cai:2008au}
H.~Cai, H.~C.~Cheng and J.~Terning,
JHEP \textbf{05}, 045 (2009)
doi:10.1088/1126-6708/2009/05/045
[arXiv:0812.0843 [hep-ph]].

\bibitem{Craig:2014aea}
N.~Craig, S.~Knapen and P.~Longhi,
Phys. Rev. Lett. \textbf{114}, no.6, 061803 (2015)
doi:10.1103/PhysRevLett.114.061803
[arXiv:1410.6808 [hep-ph]].

\bibitem{Csaki:2017jby}
C.~Cs{\'a}ki, T.~Ma and J.~Shu,
Phys. Rev. Lett. \textbf{121}, no.23, 231801 (2018)
doi:10.1103/PhysRevLett.121.231801
[arXiv:1709.08636 [hep-ph]].

\bibitem{Serra:2017poj}
J.~Serra and R.~Torre,
Phys. Rev. D \textbf{97}, no.3, 035017 (2018)
doi:10.1103/PhysRevD.97.035017
[arXiv:1709.05399 [hep-ph]].

\bibitem{Xu:2018ofw}
L.~X.~Xu, J.~H.~Yu and S.~H.~Zhu,
Phys. Rev. D \textbf{101}, no.9, 095014 (2020)
doi:10.1103/PhysRevD.101.095014
[arXiv:1810.01882 [hep-ph]].

\bibitem{Cohen:2018mgv}
T.~Cohen, N.~Craig, G.~F.~Giudice and M.~Mccullough,
JHEP \textbf{05}, 091 (2018)
doi:10.1007/JHEP05(2018)091
[arXiv:1803.03647 [hep-ph]].

\bibitem{Xu:2019xuo}
L.~X.~Xu, J.~H.~Yu and S.~H.~Zhu,
[arXiv:1905.12796 [hep-ph]].

\bibitem{Ahmed:2020hiw}
A.~Ahmed, S.~Najjari and C.~B.~Verhaaren,
JHEP \textbf{06}, 007 (2020)
doi:10.1007/JHEP06(2020)007
[arXiv:2003.08947 [hep-ph]].

\bibitem{Ade:2013sjv}
P.~A.~R.~Ade \textit{et al.} [Planck],
Astron. Astrophys. \textbf{571}, A1 (2014)
doi:10.1051/0004-6361/201321529
[arXiv:1303.5062 [astro-ph.CO]].

\bibitem{Planck:2018vyg}
N.~Aghanim \textit{et al.} [Planck],
Astron. Astrophys. \textbf{641}, A6 (2020)
[erratum: Astron. Astrophys. \textbf{652}, C4 (2021)]
doi:10.1051/0004-6361/201833910
[arXiv:1807.06209 [astro-ph.CO]].

\bibitem{ParticleDataGroup:2024cfk}
S.~Navas \textit{et al.} [Particle Data Group],
Phys. Rev. D \textbf{110}, no.3, 030001 (2024)
doi:10.1103/PhysRevD.110.030001

\bibitem{Geller:2014kta}
M.~Geller and O.~Telem,
Phys. Rev. Lett. \textbf{114}, 191801 (2015)
doi:10.1103/PhysRevLett.114.191801
[arXiv:1411.2974 [hep-ph]].

\bibitem{Csaki:2015gfd}
C.~Csaki, M.~Geller, O.~Telem and A.~Weiler,
JHEP \textbf{09}, 146 (2016)
doi:10.1007/JHEP09(2016)146
[arXiv:1512.03427 [hep-ph]].

\bibitem{Barbieri:2015lqa}
R.~Barbieri, D.~Greco, R.~Rattazzi and A.~Wulzer,
JHEP \textbf{08}, 161 (2015)
doi:10.1007/JHEP08(2015)161
[arXiv:1501.07803 [hep-ph]].

\bibitem{Ahmed:2017psb}
A.~Ahmed,
JHEP \textbf{02}, 048 (2018)
doi:10.1007/JHEP02(2018)048
[arXiv:1711.03107 [hep-ph]].

\bibitem{Borsanyi:2016ksw}
S.~Borsanyi, Z.~Fodor, J.~Guenther, K.~H.~Kampert, S.~D.~Katz, T.~Kawanai, T.~G.~Kovacs, S.~W.~Mages, A.~Pasztor and F.~Pittler, \textit{et al.}
Nature \textbf{539}, no.7627, 69-71 (2016)
doi:10.1038/nature20115
[arXiv:1606.07494 [hep-lat]].

\bibitem{vanRitbergen:1997va}
T.~van Ritbergen, J.~A.~M.~Vermaseren and S.~A.~Larin,
Phys. Lett. B \textbf{400}, 379-384 (1997)
doi:10.1016/S0370-2693(97)00370-5
[arXiv:hep-ph/9701390 [hep-ph]].

\bibitem{ParticleDataGroup:2018ovx}
M.~Tanabashi \textit{et al.} [Particle Data Group],
Phys. Rev. D \textbf{98}, no.3, 030001 (2018)
doi:10.1103/PhysRevD.98.030001

\bibitem{Zhou:2024doz}
J.~Z.~Zhou, Y.~T.~Kuang, D.~Wu, F.~Y.~Chen, H.~L{\"u} and Z.~Chang,
JCAP \textbf{12}, 021 (2024)
doi:10.1088/1475-7516/2024/12/021
[arXiv:2409.07702 [gr-qc]].

\bibitem{Kohri:2018awv}
K.~Kohri and T.~Terada,
Phys. Rev. D \textbf{97}, no.12, 123532 (2018)
doi:10.1103/PhysRevD.97.123532
[arXiv:1804.08577 [gr-qc]].

\bibitem{Lee:1956qn}
T.~D.~Lee and C.~N.~Yang,
Phys. Rev. \textbf{104}, 254-258 (1956)
doi:10.1103/PhysRev.104.254

\bibitem{Kobzarev:1966qya}
I.~Y.~Kobzarev, L.~B.~Okun and I.~Y.~Pomeranchuk,
Sov. J. Nucl. Phys. \textbf{3}, no.6, 837-841 (1966)

\bibitem{Blinnikov:1982eh}
S.~I.~Blinnikov and M.~Y.~Khlopov,
Sov. J. Nucl. Phys. \textbf{36}, 472 (1982)
ITEP-11-1982.

\bibitem{Dolgov:1998sf}
A.~D.~Dolgov, S.~H.~Hansen and D.~V.~Semikoz,
Nucl. Phys. B \textbf{543}, 269-274 (1999)
doi:10.1016/S0550-3213(98)00818-9
[arXiv:hep-ph/9805467 [hep-ph]].

\bibitem{Mangano:2001iu}
G.~Mangano, G.~Miele, S.~Pastor and M.~Peloso,
Phys. Lett. B \textbf{534}, 8-16 (2002)
doi:10.1016/S0370-2693(02)01622-2
[arXiv:astro-ph/0111408 [astro-ph]].

\bibitem{Mangano:2005cc}
G.~Mangano, G.~Miele, S.~Pastor, T.~Pinto, O.~Pisanti and P.~D.~Serpico,
Nucl. Phys. B \textbf{729}, 221-234 (2005)
doi:10.1016/j.nuclphysb.2005.09.041
[arXiv:hep-ph/0506164 [hep-ph]].

\bibitem{deSalas:2016ztq}
P.~F.~de Salas and S.~Pastor,
JCAP \textbf{07}, 051 (2016)
doi:10.1088/1475-7516/2016/07/051
[arXiv:1606.06986 [hep-ph]].

\bibitem{Hagedorn:1965st}
R.~Hagedorn,
Nuovo Cim. Suppl. \textbf{3}, 147-186 (1965)
CERN-TH-520.

\bibitem{Dashen:1969ep}
R.~Dashen, S.~K.~Ma and H.~J.~Bernstein,
Phys. Rev. \textbf{187}, 345-370 (1969)
doi:10.1103/PhysRev.187.345

\bibitem{Venugopalan:1992hy}
R.~Venugopalan and M.~Prakash,
Nucl. Phys. A \textbf{546}, 718-760 (1992)
doi:10.1016/0375-9474(92)90005-5

\bibitem{Borsanyi:2013bia}
S.~Borsanyi, Z.~Fodor, C.~Hoelbling, S.~D.~Katz, S.~Krieg and K.~K.~Szabo,
Phys. Lett. B \textbf{730}, 99-104 (2014)
doi:10.1016/j.physletb.2014.01.007
[arXiv:1309.5258 [hep-lat]].

\bibitem{Bellwied:2015lba}
R.~Bellwied, S.~Borsanyi, Z.~Fodor, S.~D.~Katz, A.~Pasztor, C.~Ratti and K.~K.~Szabo,
Phys. Rev. D \textbf{92}, no.11, 114505 (2015)
doi:10.1103/PhysRevD.92.114505
[arXiv:1507.04627 [hep-lat]].

\bibitem{Borsanyi:2010cj}
S.~Borsanyi, G.~Endrodi, Z.~Fodor, A.~Jakovac, S.~D.~Katz, S.~Krieg, C.~Ratti and K.~K.~Szabo,
JHEP \textbf{11}, 077 (2010)
doi:10.1007/JHEP11(2010)077
[arXiv:1007.2580 [hep-lat]].

\bibitem{HotQCD:2014kol}
A.~Bazavov \textit{et al.} [HotQCD],
Phys. Rev. D \textbf{90}, 094503 (2014)
doi:10.1103/PhysRevD.90.094503
[arXiv:1407.6387 [hep-lat]].

\bibitem{Ginsparg:1980ef}
P.~H.~Ginsparg,
Nucl. Phys. B \textbf{170}, 388-408 (1980)
doi:10.1016/0550-3213(80)90418-6

\bibitem{Braaten:1995jr}
E.~Braaten and A.~Nieto,
Phys. Rev. D \textbf{53}, 3421-3437 (1996)
doi:10.1103/PhysRevD.53.3421
[arXiv:hep-ph/9510408 [hep-ph]].

\bibitem{Appelquist:1981vg}
T.~Appelquist and R.~D.~Pisarski,
Phys. Rev. D \textbf{23}, 2305 (1981)
doi:10.1103/PhysRevD.23.2305

\bibitem{Shuryak:1977ut}
E.~V.~Shuryak,
Sov. Phys. JETP \textbf{47}, 212-219 (1978)
IYF-77-34.

\bibitem{Chin:1978gj}
S.~A.~Chin,
Phys. Lett. B \textbf{78}, 552-555 (1978)
doi:10.1016/0370-2693(78)90637-8

\bibitem{Kapusta:1979fh}
J.~I.~Kapusta,
Nucl. Phys. B \textbf{148}, 461-498 (1979)
doi:10.1016/0550-3213(79)90146-9

\bibitem{Toimela:1982hv}
T.~Toimela,
Phys. Lett. B \textbf{124}, 407-409 (1983)
doi:10.1016/0370-2693(83)91484-3

\bibitem{Arnold:1994ps}
P.~B.~Arnold and C.~X.~Zhai,
Phys. Rev. D \textbf{50}, 7603-7623 (1994)
doi:10.1103/PhysRevD.50.7603
[arXiv:hep-ph/9408276 [hep-ph]].

\bibitem{Arnold:1994eb}
P.~B.~Arnold and C.~x.~Zhai,
Phys. Rev. D \textbf{51}, 1906-1918 (1995)
doi:10.1103/PhysRevD.51.1906
[arXiv:hep-ph/9410360 [hep-ph]].

\bibitem{Zhai:1995ac}
C.~x.~Zhai and B.~M.~Kastening,
Phys. Rev. D \textbf{52}, 7232-7246 (1995)
doi:10.1103/PhysRevD.52.7232
[arXiv:hep-ph/9507380 [hep-ph]].

\bibitem{Kajantie:2002wa}
K.~Kajantie, M.~Laine, K.~Rummukainen and Y.~Schroder,
Phys. Rev. D \textbf{67}, 105008 (2003)
doi:10.1103/PhysRevD.67.105008
[arXiv:hep-ph/0211321 [hep-ph]].

\bibitem{Ellis:2023oxs}
J.~Ellis, M.~Fairbairn, G.~Franciolini, G.~H{\"u}tsi, A.~Iovino, M.~Lewicki, M.~Raidal, J.~Urrutia, V.~Vaskonen and H.~Veerm{\"a}e,
Phys. Rev. D \textbf{109}, no.2, 023522 (2024)
doi:10.1103/PhysRevD.109.023522
[arXiv:2308.08546 [astro-ph.CO]].

\bibitem{Linde:1980ts}
A.~D.~Linde,
Phys. Lett. B \textbf{96}, 289-292 (1980)
doi:10.1016/0370-2693(80)90769-8

\bibitem{Gross:1980br}
D.~J.~Gross, R.~D.~Pisarski and L.~G.~Yaffe,
Rev. Mod. Phys. \textbf{53}, 43 (1981)
doi:10.1103/RevModPhys.53.43

\bibitem{Laine:2006cp}
M.~Laine and Y.~Schroder,
Phys. Rev. D \textbf{73}, 085009 (2006)
doi:10.1103/PhysRevD.73.085009
[arXiv:hep-ph/0603048 [hep-ph]].

\bibitem{Davoudiasl:2004gf}
H.~Davoudiasl, R.~Kitano, G.~D.~Kribs, H.~Murayama and P.~J.~Steinhardt,
Phys. Rev. Lett. \textbf{93}, 201301 (2004)
doi:10.1103/PhysRevLett.93.201301
[arXiv:hep-ph/0403019 [hep-ph]].
\end{thebibliography}
\end{document}